\documentclass{mn2e}
\usepackage{epsfig,amssymb} 

\def\LCDM{{$\Lambda$CDM}}
\def\bj{{$\rm b_{\rm J}$}} 
\def\rf{{$\rm r_{\rm F}$}} 
\def\Mbj{{$\rm M_{\rm b_{\rm J}}^{\rm h}$}} 
\def\Mbjh{{$\rm M_{\rm b_{\rm J}} - 5 \log_{10} h$}}  
\def\Mpc{{\,h^{-1}\,{\rm Mpc}}} 
 
\def\lstar{{L$_{\star}$}}
\def\etal{{et~al.}}

\newcommand{\plotone}[1] 
           {\centering \leavevmode \psfig{file=#1,width=\columnwidth,clip=}}

\def\beq{\begin{equation}} 
\def\eeq{\end{equation}} 
\def\bea{\begin{eqnarray}} 
\def\eea{\end{eqnarray}} 
 
\def\simlt{\lower.5ex\hbox{$\; \buildrel < \over \sim \;$}}
\def\simgt{\lower.5ex\hbox{$\; \buildrel > \over \sim \;$}}

\title[The 3-point galaxy correlation function measured from the 2dFGRS]
{Statistical Analysis of Galaxy Surveys-II. 
The 3-point galaxy correlation function measured from the 2dFGRS.} 
 
\author[Gazta\~naga  et~al.]{ 
E. Gazta\~{n}aga$^{1}$, 
P. Norberg$^{2}$, 
C. M. Baugh$^3$,   
D. J. Croton$^{4}$\\ 
$^1$ Instituto de Ciencias del Espacio (IEEC/CSIC),
F. de Ciencies UAB, Torre C5- Par- 2a, Bellaterra (08193 BARCELONA) \\
$^2$ETHZ Institut f\"ur Astronomie, HPF G3.1, ETH H\"onggerberg, CH-8093 
       Z\"urich, Switzerland. \\ 
$^3$Department of Physics, University of Durham, South Road,  
    Durham DH1 3LE, UK.  \\
$^4$Max-Planck-Institut f\"ur Astrophysik, D-85740 Garching, Germany. \\ 
} 
 
\date{Accepted ---. Received ---;in original form ---} 
 
\pubyear{2005} 
 
\begin{document} 
 
\maketitle 
 
%%\label{firstpage} 
 
\begin{abstract} 
We present new results for the 3-point correlation function, $\zeta$,
measured as a function of scale, luminosity and colour from the final
version of the two-degree field galaxy redshift survey (2dFGRS).  The
reduced three point correlation function, $Q_3 \sim \zeta/\xi^2$, is
estimated for different triangle shapes and sizes, employing a full
covariance analysis. The form of $Q_3$ is consistent with the
expectations for the $\Lambda$-cold dark matter model, confirming that
the primary influence shaping the distribution of galaxies is
gravitational instability acting on Gaussian primordial fluctuations.
However, we find a clear offset in amplitude between $Q_3$ for
galaxies and the predictions for the dark matter. We are able to rule
out the scenario in which galaxies are unbiased tracers of the mass at
the 9-$\sigma$ level. On weakly non-linear scales, we can interpret
our results in terms of galaxy bias parameters.  We find a linear bias
term that is consistent with unity, $b_1 = 0.93^{+0.10}_{-0.08}$
and a quadratic bias $c_2 = b_2 /b_1 = -0.34^{+0.11}_{-0.08}$.
This is the first significant detection of a non-zero quadratic bias,
indicating a small but important non-gravitational contribution to the
three point function.  Our estimate of the linear bias from the three
point function is independent of the normalisation of underlying
density fluctuations, so we can combine this with the measurement of
the power spectrum of 2dFGRS galaxies to constrain the amplitude of
matter fluctuations.  We find that the {\it rms} linear theory
variance in spheres of radius $8h^{-1}$Mpc 
is $\sigma_8 = 0.88^{+0.12}_{-0.10}$, providing an independent 
confirmation of values derived from other techniques.  
On non-linear scales, where $\xi>1$, we find that $Q_3$ has a 
strong dependence on scale, colour and luminosity.
\end{abstract} 
                                   
 
\section{Introduction} 
 
The higher order statistics of galaxy clustering encode fundamental
information about two key dynamical aspects of the large scale
structure of the Universe: the growth mechanism of fluctuations and
the connection between the galaxy distribution and the underlying mass
(for a review, see Bernardeau et~al. 2002).  An accurate measurement
of the three-point correlation function of galaxies has the potential
to test the gravitational instability paradigm of structure formation
and, on scales that are evolving in the weakly non-linear regime, to
separate the effects of gravity from the contributions arising from
galaxy bias (Fry \& Gazta\~naga 1993; Frieman \& Gazta\~naga 1994).
 
The measurement of the three-point function and other higher order
statistics from galaxy catalogues has a rich history (Peebles \& Groth
1975; Groth \& Peebles 1977;
Fry \& Peebles 1978; Baumgart \& Fry 1991; Gazta\~naga 1992;
Bouchet et~al. 1993; Fry \& Gazta\~naga 1994).  In the past decade,
three-point statistics have supported the basic premise of
gravitational instability from Gaussian initial conditions (Frieman \&
Gazta\~naga 1994; Jing \& B\"orner 1998; Frieman \& Gazta\~naga 1999;
Hoyle, Szapudi \& Baugh 2000; Feldman et~al. 2001).  The impact of
these measurements on theoretical models has, however, not been as
great as it could have been for two reasons.  First, the traditional
theoretical predictions rely upon the application of perturbation
theory, which limits the comparison with data to relatively large
scales on which the fluctuations are evolving in a linear or 
weakly non-linear fashion.
Second, previous generations of galaxy surveys simply covered too
little volume to permit accurate measurements of the higher order 
correlation functions on the scales that
could strongly constrain the simple theoretical models.

Recent theoretical and observational advancess have been such that
we are now in a position to realize the full potential of
higher order statistics.  Theoretical models of galaxy formation have
progressed sufficiently to make predictions for the number of
galaxies that reside in dark matter haloes of different mass (Benson
et~al. 2000; Peacock \& Smith 2000; Scoccimarro et~al. 2001b; 
Berlind et~al. 2004).  This 
allows the prediction to be extended to scales for which perturbation 
theory is not valid, and provides a framework for testing the physics of galaxy
formation directly against clustering measurements.  Observationally,
two recent surveys have revolutionized our view of the local Universe:
the two-degree field galaxy redshift survey (hereafter 2dFGRS; Colless
et~al. 2001) and the Sloan Digital Sky Survey (SDSS; York 
et~al. 2000).  The ten-fold increase in survey size achieved by these
projects means that precision measurements of higher order statistics
are now be possible across a range of scales (Matarrese, Verde \& Heavens 1997;
Colombi, Szapudi \& Szalay 1998; Szapudi, Colombi \& Bernardeau 1999; 
Scoccimarro, Sefusatti \& Zaldarriaga 2004;
Sefusatti \& Scoccimarro 2005).  The higher order clustering
measurements that are possible with these surveys have the potential
to tighten the accepted values of basic cosmological parameters and to
constrain the physics of galaxy formation that govern how galaxies are
clustered.

There have have been several analyses of the distribution of
counts-in-cells using the final 2dFGRS catalogue. Baugh \etal\ (2004)
demonstrated that the higher order correlation functions display a
hierarchical scaling, $S_{p} = \bar{\xi}_{p}/\bar{\xi}^{p-1}_{2}$, where 
$\bar{\xi_{p}}$ is the $p$-point, volume averaged correlation function; 
this behaviour is expected if gravity plays a dominant role in
shaping the distribution of galaxies.  Croton \etal\ (2004b) found that
the scaling of the hierarchical co-efficients show a weak (if any)
dependence on galaxy luminosity. In the case of the three point volume
averaged correlation function, both authors found that the skewness,
$S_{3} = \bar{\xi}_{3}/\bar{\xi}^{3}_{2} \simeq 2$.  This value was
found to be independent of cell size, though both Baugh \etal\ and
Croton \etal\ noted that two large superstructures in the 2dFGRS
volume broke this scale invariance in catalogues characterised by
\lstar\ galaxies.  The result for the skewness of galaxies, $S_3
\simeq 2$, is at odds with the expectation for a \LCDM\ cosmology, in
which $S_3^{\tt DM} \simeq 3$.  
We note that Conway \etal\ (2005) and Wild \etal\ (2005) have
also looked at the constraints that the distribution of
counts-in-cells in the 2dFGRS provide on galaxy bias.  
All results from the 2dFGRS are in line with most previous 
measurements of the skewness and 3-point statistics, which 
are generally lower than the
\LCDM~ predictions (for a review, see \S8 in Bernardeau \etal\ 2002).  
This posses a puzzle, because the corresponding
measurements for the variance and the 2-point function seem to
follow the unbiased \LCDM~ predictions closely on large scales.
%We plan to address this problem in two papers. In the first paper of this 
%series (Norberg et~al. 2005;  Paper I) we will present a study of the 2-point 
%function, while the current paper is devoted to the 3-point function.

In this paper, we present the first general results for the
three-point correlation function measured from the final 2dFGRS
catalogue. Preliminary measurements of three-point statistics were
made using early releases of the 2dFGRS and SDSS datasets by Verde et
al. (2002), Jing \& B\"orner (2004), Wang et al. (2004) and Kayo
et~al. (2004). Pan \& Szapudi (2005) measured the monopole moment of
the three point function in the completed 2dFGRS. Our analysis has the
advantage over that of Pan \& Szapudi in that it includes information
about the shapes of the triangles of galaxies.  A further improvement
over previous approaches is a proper treatment of the correlation
between data points. We follow the methodology introduced by
Gazta\~naga \& Scoccimarro (2005, hereafter GS05) to obtain
constraints on bias parameters from measurements of the reduced
three-point correlation function.

This paper is organized as follows. In Section~\ref{sec:3pt}, we
review some basic definitions involving the three-point function, as
well as the methodology used for its estimation.  In
Section~\ref{sec:2dfdata}, we present the 2dFGRS catalogues and the
associated mocks.  Our results are presented in
Section~\ref{sec:q3}. This is quite a long section that is divided
into many subsections; a detailed route map is provided at the start
of this section.  Our results are compared with previous analyses of
the 2dFGRS in Section~5.  Finally, our conclusions are presented in
Section~6.
 
\section{Theoretical background} 
\label{sec:3pt}
 
In this section, we first give some basic definitions (\S2.1), before
discussing the expected form of the three point correlation function
(\S2.2). We then explain how our results can be related to the
predictions for the three point function of dark matter (\S2.3).  For
a comprehensive discussion of this material, we refer the reader to
the review by Bernardeau et~al. (2002).  The method for estimating the
three point function for the 2dFGRS is set out in \S~\ref{ssec:est}.
Finally, in \S~\ref{ssec:svd}, we give an outline of how our measurement
of the three point correlation function can be used to place
constraints on models of bias (for a complete discussion see GS05).

\subsection{Basic definitions: triangle shape and scale} 
 
\begin{figure} 
\plotone{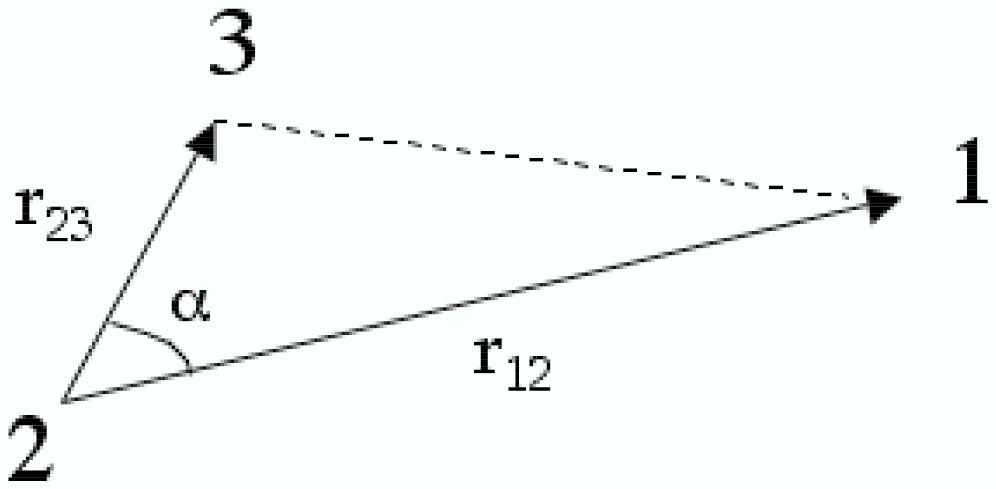} 
\vskip -5.5cm
\caption{3-points define a triangle, which is characterized
here by the two sides $r_{12}$ and $r_{13}$ and the interior angle $\alpha$.}
\label{fig:3pt} 
\end{figure}

GS05 discuss the merits of various conventions for defining triangle
shapes and scales. We adopt their preferred scheme in which a triangle
is defined by the ratio of the lengths of two of the sides of the
triangle, $\vec r_{12}/\vec r_{23}$ and the angle between them,
$\alpha$:
\beq 
\cos(\alpha)= {\vec r_{12}\over{r_{12}}} . {\vec r_{23}\over{r_{23}}}. 
\eeq 
The angle $\alpha$ can vary between $0-180$ degrees; for $\alpha=0$,
the third side of the triangle is given by $r_{31} = r_{12} - r_{23}$
and for $\alpha=180$ degrees, $ r_{31} = r_{12} + r_{23}$
(Fig.~\ref{fig:3pt}).
 
%\subsection{The form of the three point correlation function}   
\subsection{The three point correlation function}   
 
The connected two and three point correlation functions are defined as: 
\bea 
\xi(r_{12}) &=& < \delta(r_1) \delta(r_2) > \\ 
\zeta(r_{12},r_{23},r_{13}) &=& < \delta(r_1) \delta(r_2) \delta(r_3) >  
\eea 
where $\delta(r) = \rho(r)/\bar{\rho}-1$ is the local density
fluctuation around the mean $\bar{\rho}=<\rho>$ and the expectation
value is taken over different realizations of the model or physical
process.  In practice, the expectation value is constructed by
averaging over different spatial locations in the Universe, which are
assumed to form a fair sample (Peebles 1980).
 
The two and three point correlation functions change rapidly in
amplitude as a function of separation. In order to study the
relationship between the correlation functions in more detail, it is
useful to define the reduced three point function, $Q_3$, (Groth \&
Peebles 1977):
\begin{eqnarray} 
Q_3 &=& \frac{\zeta(r_{12},r_{23},r_{13})}{\zeta_H(r_{12},r_{23},r_{13})} 
\label{fourtheq} \\ 
\zeta_H  &\equiv&  
{\xi (r_{12})\xi(r_{23})+\xi(r_{12})\xi(r_{13})+\xi(r_{23})\xi(r_{13})}. 
\label{fiftheq} 
\end{eqnarray} 
\noindent 
Here we have introduced a `hierarchical' form for the three point
function, $\zeta_H$. This quantity is built up from the products of
two-point functions generated from cyclic permutations of the pair
separations which make up the sides of the triangle.  When $Q_3$ is
constant the dependence of the three point correlation function on
triangle shape and scale is fully accounted for by the corresponding
changes in $\zeta_H$; in this case $Q_3$ is said to have no
configuration dependence.
 
Previously, $Q_3$ was thought to be approximately constant as a
function of triangle size or shape (ie see Groth \& Peebles 1977), 
a phenomenon that is usually
referred to as hierarchical scaling \cite{PB80}.  However, GS05 showed
that with sufficiently accurate theoretical predictions or for
carefully constructed measurements, $Q_3$ is not in fact constant in
{\it any} clustering regime. Nevertheless, the variation in $Q_3$ with
scale is small when compared to the corresponding changes in $\xi$ or
$\zeta_H$. On small scales ($< 10~\Mpc$), and for galaxies in redshift
space (i.e. as measured by galaxy redshift surveys), GS05 showed that
$Q_3(\alpha)$ displays a characteristic U-shape anisotropy moving from
collapsed or elongated ($\alpha \sim 0, 180$) to more open ($\alpha
\sim 90$) triangles.  This effect is driven by the velocity dispersion
of galaxies inside virialized structures. GS05 demonstrated that this
U-shape is universal, being only very weakly dependent on scale, the
primordial spectral index, or the values of the cosmological
parameters. GS05 further demonstrated that this feature should be
detectable in current galaxy surveys, even if the measurements are
affected by shot-noise or if galaxies are biased tracers of the mass.
On larger scales, the impact of velocity dispersion on the form of
$Q_3$ is reduced, with the consequence that the U-shape
tends towards more of a V-shape and approaches the (real space)
perturbation theory prediction (see fig.~2 in GS05).
 
%The weak dependence of $Q_3$ on triangle shape, can be understood in terms 
%of the behaviour of $\zeta_H$ as the opening angle of the triangle, $\alpha$, 
%varies. It turns out that $\zeta_H(\alpha)$ is a slowly varying and 
%monotonically decreasing function of $\alpha$. If we consider the simple 
%case where $r_{12}$ and $r_{13}$ are held fixed, then we can write $\zeta_H$ 
%as $\zeta_H = A+B\xi(r_{23})$, where $A$ and $B$ are constants 
%($A=\xi_{12}\xi_{13}$ and $B=\xi_{12}+\xi_{23}$). The third side of the 
%triangle, $r_{23}$, changes in this case from $r_{23}=r_{13}-r_{23}$, 
%when $\alpha =0$, to  $r_{23}=r_{13}+r_{23}$ for $\alpha = 180$. Typically, 
%$\zeta_H(\alpha)$ changes by only a factor of two between $\alpha=0$ and 
%$\alpha=180$ under these conditions, thus explaining the weak dependence 
%of $Q_3$ on $\alpha$.  
 
%\subsection{Theoretical interpretation: comparison with predictions for the dark matter} 
\subsection{Theoretical interpretation} 
\label{sec:theory}
 
In order to interpret our measurements of $Q_3$ for galaxies, we will
compare them with theoretical predictions for dark matter in a
$\Lambda$CDM universe. We first explain the form expected for the
three point function of dark matter (\S2.3.1), before introducing a
notation to quantify the differences found between galaxy and dark
matter $Q_3$ measurements (\S2.3.2).

\subsubsection{The three point function of dark matter}

We shall denote the value of $Q_3$ for the dark matter by $Q^{\tt
DM}_{3}$.  The theoretical predictions for $Q^{\tt DM}_{3}$ are
relatively insensitive to the precise values of the cosmological
parameters, but have a strong dependence on the local spectral index,
$n$, of the linear perturbation theory power spectrum, $P(k)$, where
$n = {\rm d}\log P/{\rm d}\log k$ (eg. see fig.~9 and fig.~10 in
Bernardeau et~al. 2002).  This is also the case in redshift space (see
fig.~4 of GS05), but where the dependence is however weaker.
%% This next sentence is true but contradicts the sentences above
%In $\Lambda$CDM cosmologies, the spectral index $n$ is the sum of the 
%primordial index $n_s$ and the logarithmic slope of the square of the 
%transfer function, which does depend upon the cosmological parameters.
On the scales of interest to the present work, a change in the local
spectral index of $\Delta n$ translates roughly to a change in the
mean amplitude of $Q^{\tt DM}_{3}$ by $\Delta Q_3 \simeq \Delta S_3/3
\simeq \Delta n/3$ (see Juszkiewicz, Bouchet \& Colombi 1993).  As an
illustration, the difference in the shape of the power spectrum
between CDM models with density parameters of $\Omega_m=0.7$ and
$\Omega_m=0.2$ is approximately $\Delta n \simeq 0.6$ on the scales of
interest here, and so the change in $Q^{\tt DM}_{3}$ between these
models is small, $\Delta Q_3 \simeq 0.2$ (in good agreement with
fig.~4 of GS05).  The relative insensitivity of $Q^{\tt DM}_{3}$ to
changes in the CDM power spectrum is important as it strengthens any
conclusions we reach about differences between the value of $Q_3$
measured for galaxies and the predictions for the dark matter.  The
current levels of uncertainty on the matter density parameter,
$\Omega_{m}$ and the primordial spectral index $n_{s}$ are around the
10\% level or better, and so the predicted value of $Q^{\tt DM}_{3}$
is tightly constrained (e.g. Percival et~al. 2002; Tegmark
et~al. 2004; Sanchez et~al. 2005, in prep.).

It is also possible to use an empirical approach to estimate $Q^{\tt
  DM}_{3}$, without appealing directly to the $\Lambda$CDM model.  If
we assume that the two-point function of galaxy clustering has the
same shape as that of the underlying mass, then the measured
correlation function or power spectrum of galaxies can be used to
infer the spectral index of the dark matter.  The two-point
correlation function and the power spectrum of galaxy clustering have
both been measured for the 2dFGRS on large scales (Percival
et~al. 2001; Hawkins et~al. 2003; Cole et~al. 2005).  It turns out
that the shapes of these clustering statistics are compatible with the
predictions of the $\Lambda$CDM model.  The uncertainties in $n$ are
small ($\Delta n <0.1$) compared to the sampling errors in the
measurements of $Q_3$ for the 2dFGRS (GS05).  We will therefore assume
the concordance $\Lambda$CDM model, specified by $\Omega_m \simeq
0.3$, $\Omega_\Lambda \simeq 0.7$ and $h \simeq 0.7$, to generate
predictions for $Q^{\tt DM}_{3}$, and neglect the impact of any
uncertainty in these parameters.

\subsubsection{Comparing the three point functions of galaxies and dark 
matter: the implications for bias}

The $Q_3$ value measured for galaxies may be different from the
theoretical predictions for the dark matter, $Q^{\tt DM}_{3}$.  We
adopt a particularly simple scheme to quantify any such differences:
\beq 
Q_3 \simeq {1\over{B}} ~\left(Q_3^{\tt DM}+ C\right).  
\label{eq:Q3G} 
\eeq 
Two numbers specify the difference between the measured and predicted
$Q_3$: a shift or offset, $C$, and a scaling, $B$.  The simple ansatz
given in Eq.~\ref{eq:Q3G} is general and can, in principle, be applied
on any scale.  However, the interpretation of the numbers $B$ and $C$
does depend upon the scale under consideration. Furthermore, we should
caution that this model may not necessarily always provide a good
description of the transition from the clustering of the dark matter
to galaxies.

The form we have chosen is motivated by perturbation theory, which
applies on scales for which the correlations are small, i.e. $\xi<1$.
Fry \& Gazta\~naga (1993) modelled fluctuations in the density of
galaxies, $\delta_G$, as a local, non-linear expansion of fluctuations
in the mass distribution, $\delta$:
\beq 
\delta_G = F[\delta] \simeq \sum_{k}~{b_k\over{k!}}~\delta^k. 
\label{eq:bk} 
\eeq 
This formalism can be used to derive a relation between the three
point function of galaxies and mass (see Fry \& Gazta\~naga 1993;
Frieman \& Gazta\~naga 1994).  On the weakly non-linear scales for
which this transformation is a reasonable approximation, $B \approx
b_{1}$ and $C \approx c_2 = b_{2}/b_{1}$:
\beq 
Q_3 \simeq {1\over{b_1}} ~\left(Q_3^{\tt DM}+ c_2\right).  
\label{eq:Q3Gbk} 
\eeq   
In this case, the shift by $C$ can be interpreted as a
non-gravitational contribution to $Q_3^{\tt DM}$ and $B$ is a simple
linear bias scaling.  These effects can become degenerate if $Q_3$ is
approximately constant or when the measurement errors become large.
Nevertheless, it is possible, in principle, to compare the shape of
$Q_3$ measured for galaxies to that predicted for the dark matter,
and so constrain $B$ and $C$ separately.  Norberg et~al. (2005; Paper
I, in prep.) use the two point correlation function to obtain a working
definition of the scale marking the approximate boundary between the
non-linear and weakly non-linear regimes; they propose that weakly
non-linear scales ($\xi \ll 1$) correspond to pair separations of
$\gtrsim 9~\Mpc$, whereas the non-linear regime ($\xi > 1$) is
reached when $r \lesssim 6~\Mpc$.

From its definition in Eq.~\ref{fourtheq}, $Q_3$ is independent of the
amplitude of fluctuations on large scales.  We can therefore use the
value of $B$ to constrain the amplitude of fluctuations in the dark
matter. In this approach, we take the two-point correlation function
measured for galaxies and divide this by $B^{2}$ to obtain an
empherical two-point function estimate of the dark matter.  Then,
after measuring the {\it actual} shape of the two-point function of
the dark matter distribution from simulations, we can constrain the
{\it rms} linear variance in spheres of radius $8 h^{-1}$Mpc,
$\sigma_{8}$ by equating our empirical dark matter estimate to the
actual value.  This method for constraining $\sigma_{8}$ relies upon
several approximations and assumptions that we have tested
successfully using N-body simulations (see Norberg \etal\ 2005, Paper
I, in prep, for a full description of the method).  Similar approaches have
already been attempted using the skewness of the distribution of
galaxy counts-in-cells, $S_3$ (Fry \& Gazta\~naga 1993; Gazta\~naga
1994; Gazta\~naga \& Frieman 1994), the bispectrum (eg Frieman \&
Gazta\~naga 1994; Fry 1994; Scoccimarro 1998; Verde et~al. 2002) and
the angular 3-point function (Frieman \& Gazta\~naga 1999).

\subsection{The estimation of the three point function} 
\label{ssec:est} 

To estimate the three point correlation function efficiently for the
2dFGRS, we use the fast grid based algorithm introduced by Barriga \&
Gazta\~{n}aga (2002). GS05 presented further tests of this algorithm
using a wide range of numerical simulations and mock catalogues. These
authors demonstrated that special attention should be paid to the grid
dimension employed in order to obtain robust estimates of the three
point function in redshift space.  For practical reasons, we use a
somewhat lower than ideal pixel resolution in the estimation of the
three point function from the 2dFGRS. This results in some smoothing
of the U-shape in $Q_3(\alpha)$ for collapsed configurations (compare
fig.~5 of GS05 with our Fig.~5). As we use the same pixelization in
the analysis of the mocks and dark matter theory, this loss of
resolution does not affect our conclusions, although it could result
in slightly less than optimal constraints on $B$ and $C$.

The final 2dFGRS catalogue contains some incompleteness which is
quantified by the spectroscopic completeness mask (Norberg et
al. 2002b). The spectroscopic completeness of the final 2dFGRS is much
more uniform than that of the 100K release or the samples used in
earlier clustering analyses by the 2dFGRS team (e.g.  Verde et
al. 2002), as shown by fig.~3 of Cole et al. (2005).  We reject pixels
on the sky for which the spectroscopic completeness is less than
$50\%$. We account for the remaining incompleteness by applying a
weight to the galaxy cell density.  Further details about the 2dFGRS
catalogue are given in Section~3.1.

\subsection{Constraining model parameters using $Q_3$} 
\label{ssec:svd}

The values of $Q_3$ measured for different opening angles are
correlated.  This needs to be taken into account when using
measurements of $Q_3$ to place constraints on model parameters, such
as the values of $B$ and $C$ defined by Eq.~\ref{eq:Q3G}.  GS05
introduced an eigenmode approach to parameter fitting with $Q_3$, and
used this to demonstrate the level of the constraints on $B$ and $C$
that could be expected from the 2dFGRS. GS05 estimated the covariance
matrix for $Q_3(\alpha)$ using the mock 2dFGRS catalogues that we
describe in Section~\ref{mock}. They then obtained the inverse of the
covariance matrix using the Singular Value Decomposition method. In
this approach, eigenmodes that fall below some specified
signal-to-noise (S/N) threshold are discarded. The likelihood contours
in the $B-C$ plane are specified by $\delta \chi^2$ computed using the
eigenvectors above the S/N threshold. The S/N values that we estimate
are not quite optimal, because we use a finite number of mock
catalogues. Our errors are therefore conservative estimates.  The S/N
values indicate the significance of the measurement of $Q_3$ (i.e. the
number of standard deviations that the signal is above the
noise). However, the S/N ratio does not translate directly into the
size of the likelihood contours in the $B-C$ plane, because the
degeneracy between these parameters also depends on how far the
measured $Q_3$ deviates from a constant as a function of angle.  Even
in the case of an infinite S/N, $B$ and $C$ will be degenerate if
$Q_3$ is independent of angle (i.e. see Eq.~\ref{eq:Q3G}).

\section{The galaxy catalogues}
\label{sec:2dfdata} 

In this section, we describe the 2dFGRS data that we use to measure
$Q_3$ (\S3.1), and the synthetic catalogues that are employed to
perform our error analysis and make the $\Lambda$CDM predictions
(\S3.2).
 
\subsection{The 2dFGRS data}  
 
Our starting point is the final 2dFGRS catalogue (Colless et~al. 2003;
a full description of the construction of the survey is given in
Colless et~al. 2001).  The 2dFGRS consists of 221,414 unique, high
quality galaxy redshifts, with a median redshift of $z\approx0.11$ to
the nominal extinction corrected magnitude limit of $b_{\rm J} \approx
19.45$.  
Colour information is now available for the 2dFGRS through the addition 
of \rf-band photometry (see Cole et~al. 2005).
In our analysis, we consider the two contiguous regions of
the survey which lie towards the directions of the SGP and NGP,
covering a solid angle of approximately 1200 square degrees.  The
redshift completeness of the survey varies with position on the sky.
Colless et~al. (2001) and Norberg et~al. (2002b) describe a strategy
for dealing with this incompleteness in clustering studies.  We
restrict our attention to regions of the survey for which the
spectroscopic completeness exceeds $50$~\%. We note that the typical
completeness for the final survey is much higher than this ($\sim
85$~\%).
  
We follow the approach adopted in several previous clustering analyses
of the 2dFGRS and construct volume limited samples from the magnitude
limited redshift catalogue (Norberg et~al. 2001; Norberg et~al. 2002a;
Baugh et~al. 2004; Croton et~al. 2004a,b). This greatly simplifies the
estimation of the clustering signal, as then the only variations in
number density across the galaxy sample will be due to the presence of
large scale structure. A volume limited sample is defined by an
interval in absolute magnitude, and this translates into a minimum and
maximum redshift. For each galaxy an absolute magnitude is computed
using its redshift and apparent magnitude, and assuming the band-shift
and evolutionary correction ($k+e$) advocated by Norberg
et~al. (2002b) (see also Cole et~al. 2005).  In a volume limited
sample, each galaxy could in principle be displaced to any depth
within the sample and would still remain within the apparent magnitude
range of the survey.  In this paper we consider the samples listed in
Table~\ref{tab1}, which correspond to samples 1-4 as listed in Table 1
of Croton et~al. (2004b). As in Cole et~al. (2005), we also split the
samples by restframe \bj-\rf\ colour into blue, \bj-\rf$<$1.07, and
red, \bj-\rf$>$1.07, subsamples.

Baugh et~al. (2004) and Croton et~al. (2004b) both point out the
impact of two superstructures, one in each of the NGP and SGP regions,
on the estimation of the moments of the distribution of
counts-in-cells from the 2dFGRS. The SGP structure is at $\alpha \sim
13{\rm hr}$ and $d \simeq 240 h^{-1}$Mpc and the NGP structure is at $\alpha
\sim 0.5 {\rm hr}$ and $d \simeq 325 h^{-1}$Mpc (see fig.~1 of Baugh
et~al. 2004).  They found these overdensities were particularly
influentual on measurements made from the \lstar\ volume limited
sample, i.e. for galaxies with $-20<$\Mbjh$<-19$; fainter volume
limited samples do not extend to the distance of the superstructures
and brighter samples cover a larger volume and thus dilute the
contribution of the structures.  In this paper, we follow the approach
of these authors and in Section~\ref{sec:q3_lstar} present
measurements of the three point function made when masking out the
regions containing these superstructures. It turns out that this
exclusion removes only a small fraction of the total \lstar\ volume,
approximatly $2~\%$, along with the fewer than $5~\%$ of the total
galaxies contained within it. This exercise is merely intended to be
illustrative.  We are not proposing that the removal of these
structures should be thought of as a correction to our measurements,
but rather should serve as an indication of the magnitude of
systematic effects in the estimation of higher order statistics from
surveys of the size of the 2dFGRS.
 
\begin{table}
  \centering
  \footnotesize
  \caption{Properties of the combined 2dFGRS SGP and NGP
    volume-limited catalogues (VLCs).  Columns 1 and 2 give the faint
    and bright absolute magnitudes that define the sample.  Columns 3
    and 4 give the number of galaxies in each sample and the mean
    number density. Columns 5 and 6 state the minimum and maximum
    comoving distances that bound each sample for the nominal apparent
    magnitude limits of the survey.  All distances are comoving and
    are calculated assuming standard values for the cosmological parameters
    ($\Omega_{0}=0.3$ and $\Omega_{\Lambda}=0.7$).
  }
  \label{tab1}
  \begin{tabular}{ccccrrrr}
    \hline \hline
    \multicolumn{2}{c}{Mag. range} &            {N$_{\rm G}$} &
           {$\rho_{ave}$}  &           {D$_{min}$} & {D$_{max}$}\\
    \multicolumn{2}{c}{ \Mbjh} &
         &
           {\tiny $10^{-3}/h^{-3}$Mpc$^3$} &
           {\tiny $h^{-1}$Mpc} & {\tiny $h^{-1}$Mpc} \\
           \hline
      -17.0 & -18.0 & \hphantom{0}8038 & 10.9  & 24.8 & 169.9 \\
      -18.0 & -19.0 & 23290 & \hphantom{00}9.26  & 39.0 & 255.6 \\
      -19.0 & -20.0  & 44931 & \hphantom{00}5.64& 61.1 & 375.6 \\
      -20.0 & -21.0 & 33997 & \hphantom{00}1.46   & 95.1 & 537.2\\
      -21.0 & -22.0 & \hphantom{0}6895 & \hphantom{00}0.11   & 146.4 & 747.9\\
         \hline \hline
  \end{tabular}
\end{table}

\subsection{Mock catalogues} 
\label{mock}
 
Mock catalogues play an important role in our analysis. They are used
to compute errors on our measurements and also as a means of
generating the predictions of the $\Lambda$CDM model, taking into
account the selection function of the 2dFGRS.  Following GS05, we
construct the (normalized) covariance matrix for the measurements of
the three point function using an ensemble of 22 synthetic 2dFGRS
catalogues extracted from the $\Lambda$CDM Hubble Volume N-body
simulation (Evrard et~al. 2002). The construction of these catalogues
is described by Norberg et~al. (2002b). The mock catalogues have the
same radial and angular selection function as the 2dFGRS, and have
been convolved with the completeness mask of the survey. A simple
phenomenological prescription has been applied to the final density
field in the simulation in order to extract points with a clustering
amplitude that is a modulated version of the clustering of the
underlying dark matter (Cole et~al. 1998).  We also use the dark
matter from the Hubble Volume simulation to generate dark matter
predictions for the concordance $\Lambda$CDM model in redshift space.

The mocks were not constructed to match higher order clustering
statistics, as the biasing model used (see Norberg \etal\ 2002b; Cole
\etal\ 2005) was tuned only to reproduce the 2-point
correlation function of all galaxies (as measured for the 2dFGRS by Hawkins
\etal\ 2003).  In particular, this means that the mock galaxy
catalogues do not display luminosity dependent clustering, as seen in
the data (Norberg et~al. 2001). This deficiency can be turned around
to provide an interesting test of our analysis, since the mocks should
always give, for a fixed triangle configuration, the same mean
$Q_3(\alpha)$ for different luminosities, regardless of the volume
under consideration or the density of galaxies.

\section{Results for $Q_3$} 
\label{sec:q3}

In this section we present our measurements of $Q_{3}$ for different
2dFGRS galaxy samples, defined by luminosity and colour.  In the first
two subsections, we consider general triangle shapes, focusing first
on scales for which we expect the clustering to be in the weakly
nonlinear regime (\S\ref{sec:q3_wnl}) before considering the nonlinear
regime (\S\ref{sec:q3_nl}). The reason behind this split is that the
interpretation of our measurements is quite different in these two
cases, as discussed in section~\ref{sec:theory}. In both sections, we
consider how our measurements depend upon galaxy luminosity and, in
the case of the nonlinear regime, on colour as well.
In~\S\ref{sec:q3_scale}, we consider the special case of equilateral
and elongated triangles, which give a cleaner measure of the physical
scale dependence of $Q_3$.  Finally, in~\S\ref{sec:q3_lstar}, we
discuss the influence of large structures on the measurement of the
three point function in the 2dFGRS. From now on, we use the shorthand
notation \Mbj\ to denote \Mbjh.

\begin{figure*}
\centering{{\epsfxsize=0.95\textwidth \epsfbox{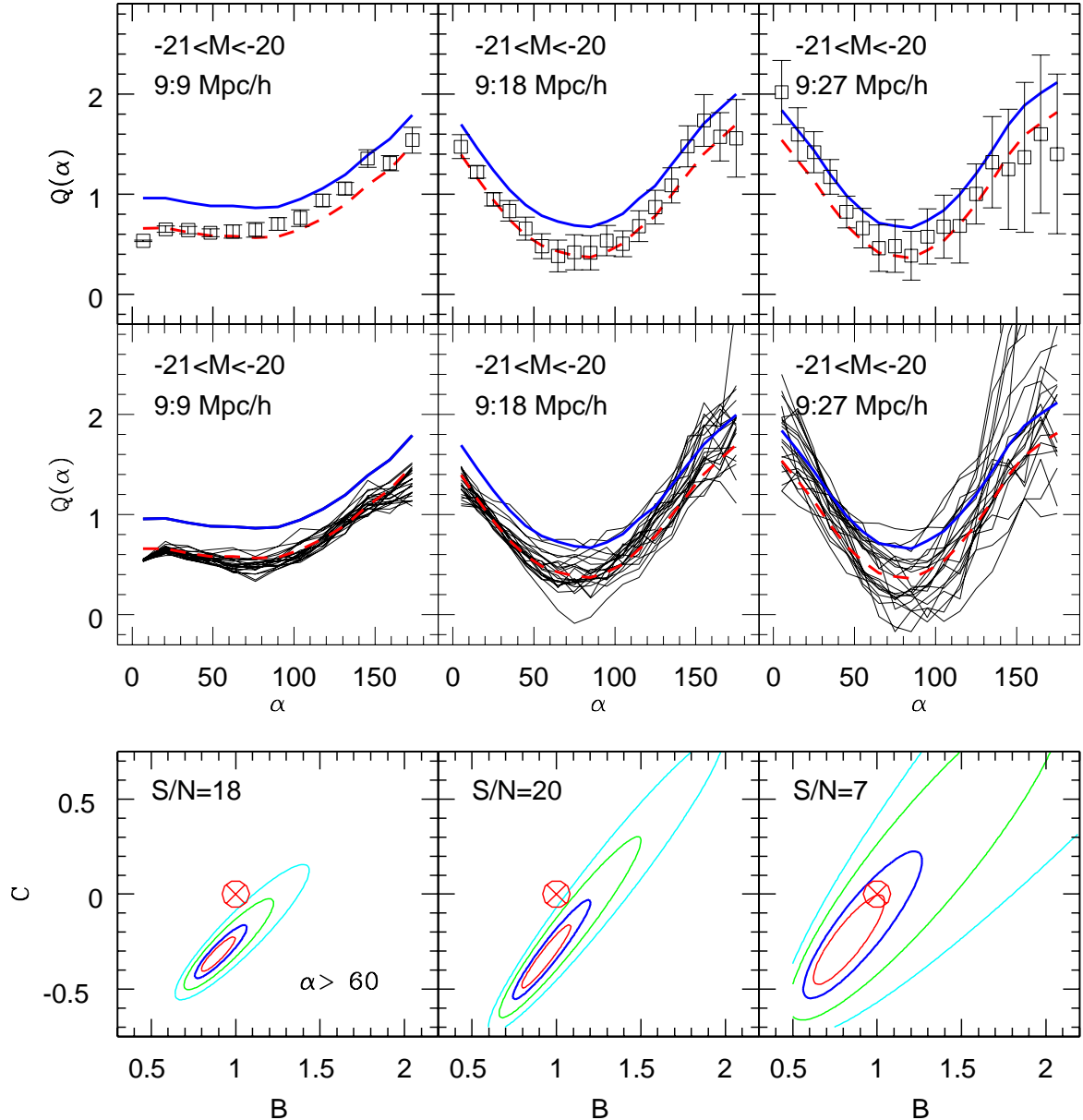}}}
\caption{$Q_3$ in the weakly nonlinear regime. The upper row shows
  measurements from the 2dFGRS $-21<$\Mbj$<-20$ volume limited
  sample. Different columns show the results for different triangle
  sizes, as indicated by the legend.  The squares show the mean value
  of $Q_3$ as a function of $\alpha$ and the error bars are derived
  from the scatter in the mock catalogues with the same magnitude
  limits. The thick solid curves in the upper two rows show the
  predictions for $Q_3$ in the $\Lambda$CDM model. The thick dashed
  curves in these panels shows the effect of applying a transformation
  (Eq.~\ref{eq:Q3G}) to this prediction corresponding to $B=1$ and
  $C=-0.3$.  The thin solid lines in the middle row show the mean
  $Q_3$ measured in individual mocks. In the bottom row, we show the
  constraints on $B$ and $C$, derived from an eigenmode analysis. The
  four contours shown from outside in correspond to
  $\chi^2=11,8,6.17,2.3$ (ie 99.7\%, 95.4\% and 68.3\% confidence
  interval for 2 parameters) and $\chi^2=1$ (ie 68.3\% for one of the
  parameters). The cross point shows $B=1$, $C=0$ for reference.  }
\label{fig:q3_210_wnl} 
\end{figure*}

\subsection{$Q_3(\alpha)$ in the weakly non-linear regime} 
\label{sec:q3_wnl} 

The $-21<$\Mbj$<-20$ volume limited sub-sample yields the highest
signal-to-noise measurements of $Q_3$ in the weakly non-linear regime.
Samples brighter than this have too low a galaxy density to permit
robust measurements on such scales, while fainter samples span smaller
volumes, resulting in a larger sampling variance.  We will therefore
describe the results for this sample in some detail, as this will
serve to make several basic points that can be applied to other samples.

\subsubsection{$Q_3(\alpha)$ for $-21<$\Mbj$<-20$} 
\label{sec:q3_210_wnl} 

In the top row of Fig.~\ref{fig:q3_210_wnl}, we show $Q_3(\alpha)$ for
galaxies with $-21<$\Mbj$<-20$ and different triangle configurations:
$r_{12}=9$ and $r_{13}=9$, 18 and $27~\Mpc$, from left to right
respectively.  The middle panel illustrates the scatter expected in
such a measurement of $Q_3(\alpha)$, obtained using the mock galaxy
catalogues. Here, the prediction for the $\Lambda$CDM dark matter
model is shown by the solid line while the dashed line shows a biased
version of this, computed by inserting $B=1$ and $C=-0.3$ into
Eq.~\ref{eq:Q3G}. In the bottom panels, we show the likelihood
contours for $B$ and $C$ derived from fitting the observed
$Q_3(\alpha)$ to the $\Lambda$CDM prediction with the theoretically
motivated relation given by Eq.~\ref{eq:Q3G}.
 
In the top panels of Fig.~\ref{fig:q3_210_wnl} we clearly see, for the
data, the characteristic dependence of $Q_3$ on $\alpha$, with the
V-shape becoming more pronounced as larger scales are considered. Note
how the variation in the shape of $Q_3$ with scale seen in the 2dFGRS
data is mimicked by the dark matter predictions and by results for the
mock catalogues.
 
The middle panels of Fig.~\ref{fig:q3_210_wnl} show how closely $Q_3$
estimated from the mock catalogues agrees with the measurements from
the 2dFGRS. This agreement is all the more remarkable when one recalls
that a match to $Q_3$ was not required in the construction of the
mocks.  Another noteworthy feature of the results for the mocks is the
strong covariance that is apparent between the measurements of $Q_3$
in different angular bins. Hence, to perform a meaningful fit to
$Q_3(\alpha)$, there is a clear need to decompose the measurement into
statistically independent Q-eigenmodes, yielding a basis in which the
covariance matrix is diagonal.  This strong correlation between bins
in measurements of $Q_3(\alpha)$ was originally pointed out by GS05.
%Figure~9 of GS05 shows an illustration of the joint
%covariance matrix for the configuration $r_{13}=2~r_{12}=24~\Mpc$,
%using mock galaxy catalogues and dark matter simulations
%respectively. 
Further details of the application of the singular value decomposition
to the measured values of $Q_3(\alpha)$ and the estimation of the
signal-to-noise (S/N) of the Q-eigenmodes can be found in GS05.

From Fig.~\ref{fig:q3_210_wnl}, it is clear that the $-21<$\Mbj$<-20$
sample provides a high quality measurement of $Q_3(\alpha)$ in the
weakly non-linear regime.  The characteristic V-shape dependence of
$Q_3$ on angle is readily apparent across a range of triangle scales.
With such a high S/N of the Q-eigenmode decomposition, this volume
limited sample is expected to provide strong constraints on the
parameters $B$ and $C$ in Eq.~\ref{eq:Q3G}; we recall that in the
weakly non-linear regime, $B\sim b_{1}$ and $C\sim c_{2}$.  The
constraints on these parameters, shown in the bottom row of
Fig.~\ref{fig:q3_210_wnl}, are discussed below in
section~\ref{sec:b1_c2_210_wnl}.

\subsubsection{$Q_3(\alpha)$ as function of luminosity} 
\label{sec:q3_L_wnl} 

\begin{figure} 
\centering{\epsfig{file=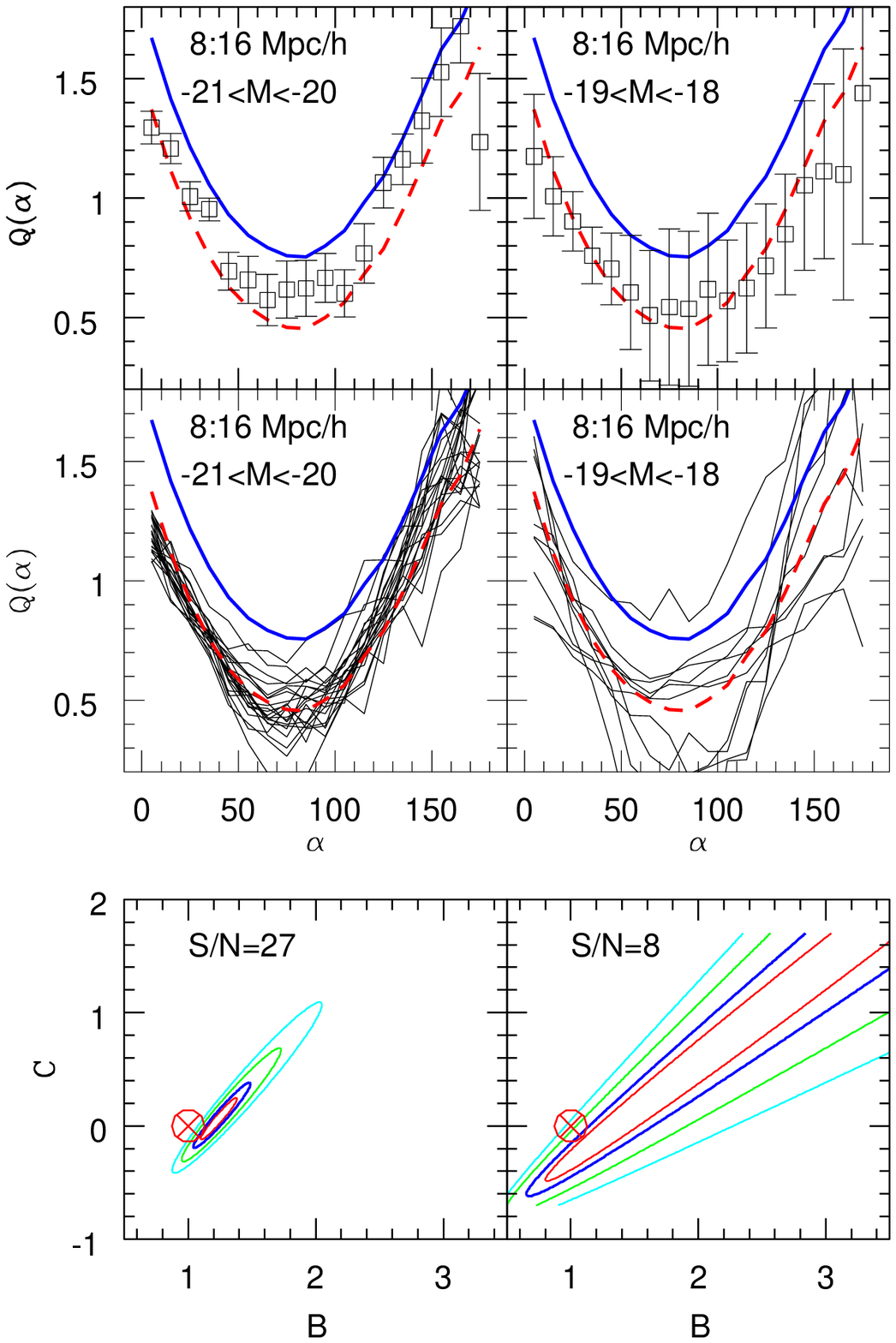,width=0.45\textwidth,clip=,bbllx=100,bblly=144,bburx=490,bbury=718}}
\caption{Same as Fig.~\ref{fig:q3_210_wnl}, but for different volume
  limited samples: $-21<$\Mbj$<-20$ (left) and $-19<$\Mbj$<-18$
  (right). In all panels, we consider the case where
  $r_{13}=2~r_{12}=16~\Mpc$.  }
\label{fig:q3_L_wnl} 
\end{figure}

Staying in the weakly non-linear regime, we now consider the
dependence of $Q_3(\alpha)$ on galaxy luminosity.  In
Fig.~\ref{fig:q3_L_wnl} we present, for the triangle configuration
$r_{13}=2~r_{12}=16~\Mpc$, the variation of $Q_3(\alpha)$ with
luminosity, as measured from volume limited catalogues defined by
$-21<$\Mbj$<-20$ and $-19<$\Mbj$<-18$.  On these weakly non-linear
scales the best S/N, as indicated in the bottom panels of
Fig.~\ref{fig:q3_L_wnl}, again occurs for the $-21<$\Mbj$<-20$ sample,
as expected from the analysis of GS05.
%% This explains our decision to focus on that sample in
%% section~\ref{sec:q3_210_wnl}. 
 
%Before continuing, astute readers will have already noticed that we
%have omitted the results of the volume limited catalogue centred on
%\lstar, covering $-20<$\Mbj$<-19$.  Indeed, for higher order
%clustering statistics in the 2dFGRS this volume limited sample rather
%special, as was pointed out in the counts-in-cells analysis of Baugh
%\etal\ (2004) and Croton \etal\ (2004). For that reason, results from
%that volume limited range are discussed in detail in a separate
%section (\S\ref{sec:q3_lstar}).
 
Fig.~\ref{fig:q3_L_wnl} shows that the characteristic $Q_3(\alpha)$
shape is seen for both bright and faint galaxies.  Moreover, the
V-shape is essentially the same in the two samples, within the
measurement errors.  Given the size of the errors, we are not yet able
to detect any clear evidence of luminosity segregation in
$Q_3(\alpha)$ in the weakly non-linear regime. Unfortunately, we
encounter the same limitation in section~\ref{sec:q3_scale}, when
considering large equilateral triangles. 

From the middle panel of Fig.~\ref{fig:q3_L_wnl}, we conclude that our
measurement of $Q_3$ is robust to sampling variance and volume
effects. Reassuringly, the results obtained from the mock catalogues
for $Q_3(\alpha)$ for different volume limited samples are, within the
errors, equivalent as they should be by construction.  The bottom
panel of Fig.~\ref{fig:q3_L_wnl}, which presents the constraints on
$B$ and $C$ (Eq.~\ref{eq:Q3G}), is discussed in the next section.

\begin{figure} 
\plotone{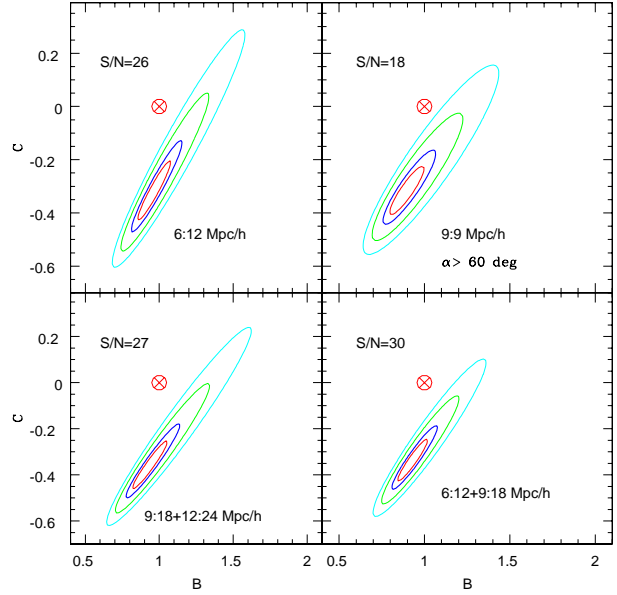} 
\caption{The combined constraints on $B$ and $C$, in the weakly
  non-linear regime, using different triangle configurations and the
  joint covariance of $Q_3(\alpha)$ for the 2dFGRS volume limited
  sample $-21<$\Mbj$<-20$.  Each panel uses different triangle
  configurations (as indicated by each legend). For isosceles
  triangles with $r_{12}=r_{13}=9~\Mpc$, we fit only for $\alpha>60$
  degrees, to ensure that only weakly non-linear scales (9 to
  $18~\Mpc$) are considered.}
\label{fig:b1_c2_210_wnl} 
\end{figure}

\begin{table} 
  \centering 
  \footnotesize 
  \caption[]{The best-fitting $b_1$ and $c_2$ values for the $-21<$\Mbj$<-20$ 
  sample for a range of weakly non-linear scales, along with the associated 
  marginalized 68~\% confidence intervals (corresponding to 1-$\sigma$ if 
  one assumes Gaussian statistics), as taken from
  Fig.~\ref{fig:b1_c2_210_wnl}. Each entry in the table uses different triangle
  configurations (indicated in brackets in the first column),  probing  
  the range of scales listed in the first column.
} 
\label{tab:q3_wnl} 
  \begin{tabular}{ccccc}  
    \hline \hline 
    scale [$\Mpc$] & $b_1$ & 68~\% C.I. & $c_2$ & 68~\% C.I. \\  \hline 
%     3-6 & $1.49$ & $[1.22, 1.81] $  & $-0.06$ & $[-0.21,+0.18]$  \\ 
     6-18 (6:12) & 1.01 & [0.91,1.15] & -0.27 & [-0.37,-0.15]  \\ 
     6-27 (6:12+9:18)& 0.93 & [0.85,1.03]  & -0.34 & [-0.42,-0.23] \\ 
     9-18 (9:9, $\alpha>60$) & 0.97 & [0.88,1.09]  & -0.31 & [-0.41,-0.24]  \\      
     9-36 (9:18+12:24) & 0.94 & [0.83,1.07]  &  -0.36 & [-0.45,-0.23]  \\ 
    \hline \hline 
  \end{tabular} 
\end{table}  

\subsubsection{Constraints on $b_1$ \& $c_2$ from $-21<$\Mbj$<-20$} 
\label{sec:b1_c2_210_wnl} 
 
The bottom panels of Figs.~\ref{fig:q3_210_wnl} and~\ref{fig:q3_L_wnl}
present the likelihood contours for the bias model parameters $B$ and
$C$, as defined in Eq.~\ref{eq:Q3G}, which relates $Q_3(\alpha)$ for
galaxies to $Q^{\tt DM}_3(\alpha)$ for the dark matter.  Remember that
in the weakly non-linear regime, $B=b_1$ and $C=c_2$, the conventional
linear and non-linear biasing terms respectively.  The cross-circle
point in the bottom panels of Figs.~\ref{fig:q3_210_wnl}
and~\ref{fig:q3_L_wnl} shows an unbiased $\Lambda$CDM prediction (i.e.
$b_1=1$ and $c_2=0$). For some configurations, such as the right
bottom panel of Fig.~\ref{fig:q3_L_wnl}, the likelihood contours are
broad and the bias parameters are poorly constrained. In this
particular case (i.e. the faint galaxy population), this is chiefly
a result of the small volume considered, telling us that these
larger-scale triangle configurations do not sample enough different
environments in this volume.  
%% Under such circumstances we can not find reliable estimates on how the
%% biasing parameter changes with luminosity on weakly non-linear scales.
 
By combining measurements on different weakly non-linear scales, we 
can improve the constraints on $b_1-c_2$ for the $-21<$\Mbj$<-20$ sample, as shown in
Fig.~\ref{fig:b1_c2_210_wnl}.  Table~\ref{tab:q3_wnl} lists the
corresponding marginalized best-fitting values. In this regime, we do
find a slight trend, with a decrease of the bias parameters with
increasing scale, although the trend is not very significant and is
within the quoted 1-$\sigma$ error. The strongest constraints on
$b_1-c_2$ come from the $6-27\,\Mpc$ configuration (bottom right panel
in Fig.\ref{fig:b1_c2_210_wnl}), with measured values of $b_1 =
0.93^{+0.10}_{-0.08}$ and $c_2 = -0.34^{+0.11}_{-0.08}$.
 
Finally, we note that in all the panels of
Fig.~\ref{fig:b1_c2_210_wnl} the unbiased $\Lambda$CDM prediction
(shown by the cross-circle) is strongly excluded by the data. For
example, for $r_{13}=2~r_{12}=18\,\Mpc$, $\Delta \chi^2 >80$ for 2
degrees of freedom (which implies a disagreement in excess of
9-$\sigma$).  Despite the correlation between $b_1$ and $c_2$, the
significance of the detection of bias, i.e. values away from $b_1=1$
and $c_2=0$, is much larger than is apparent from just adding the
errors in quadrature or using the values in Table~\ref{tab:q3_wnl}
with a square errorbox. In fact, the measured values of $b_1$ and
$c_2$ seem to follow a degenerate line, illustrated in
Fig.~\ref{fig:b1_c2_210_wnl}, that avoids $b_1=1$ and $c_2=0$ for all
scales and luminosities. As the scale or the luminosity of the sample
decreases, $b_1$ crosses unity and $c_2$ passes through zero (though
not at the same time!), following $b_1 \simeq c_2 + 1.2$ and hence
avoiding the unbiased prediction.
 
\begin{figure} 
\centering{\epsfig{file=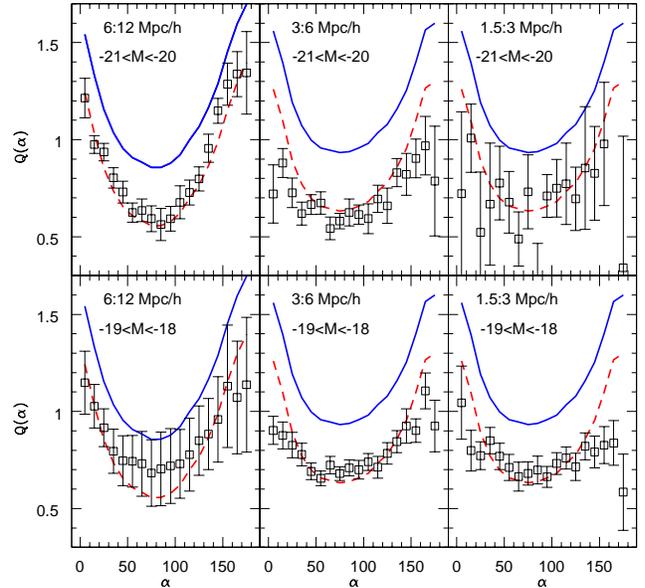,width=0.48\textwidth,clip=,bbllx=35,bblly=180,bburx=570,bbury=718}}
\caption{$Q_3(\alpha)$ in the non-linear regime for two different 
  luminosities (increasing from bottom to top) and different 
  scales (decreasing from left to right), as indicated in the 
  legend in each panel. The biased model (dashed lines) 
  has $B=1$ and $C=-0.3$, whereas $Q_3$ for
  \LCDM\, is plotted with a solid line.  }
\label{fig:q3_L_nl} 
\end{figure}

\subsection{$Q_3(\alpha)$ in the non-linear regime} 
\label{sec:q3_nl} 
 
In this section, we consider triangle configurations which probe
non-linear clustering scales corresponding to $\simlt 6\,\Mpc$ (see
section~\ref{sec:theory}).  We first discuss results for three
different triangle configurations using the $-19<$\Mbj$<-18$ volume
limited catalogue (\S\ref{sec:q3_190_nl}).  This is found to be the
optimal sample with which to study the non-linear regime due to its
relatively high galaxy number density in a volume that is large enough
to account for small-scale cosmic variance.  Thus, this sample is the
one least affected by shot noise\footnote{The \lstar\ sample is
actually statistically better than this fainter one, but suffers
from a much larger systematic uncertainty due to the presence of the
two superstructures (see section~\ref{sec:q3_lstar}).}.  Then,
in~\S\ref{sec:q3_L_nl}, we consider the variation of $Q_3(\alpha)$
with luminosity and colour on non-linear scales, followed by a study
of the constraints on the model parameters $B$ and $C$
(Eq.~\ref{eq:Q3G}) in~\S\ref{sec:B_C_nl}.
%% Unlike the previous section, we now also consider the \lstar\ $19$ to
%% $20$ volume limited sample.  Past 2dFGRS studies have shown that the
%% systematic effect of super structures, which strongly influence
%% higher-order clustering on intermediate and larger scales, do not play
%% an important role on the smaller scales that we consider here (refs).

\subsubsection{$Q_3(\alpha)$ for $-19<$\Mbj$<-18$} 
\label{sec:q3_190_nl} 
 
\begin{figure} 
\centering{\epsfig{file=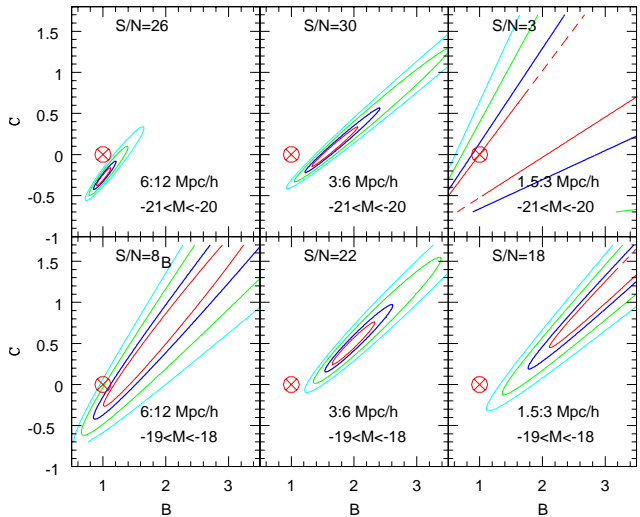,width=0.48\textwidth,clip=,bbllx=35,bblly=180,bburx=570,bbury=718}}
\caption{Constraints on the model parameters $B$ and $C$, for the
  samples presented in Fig.~\ref{fig:q3_L_nl}.}
\label{fig:B_C_L_nl} 
\end{figure}

The bottom row of Fig.~\ref{fig:q3_L_nl} shows $Q_3(\alpha)$ for the
$-19<$\Mbj$<-18$ sample with three triangle configurations which probe
non-linear scales.  From left to right, we have $r_{13}=2~r_{12}$ with
$r_{12}=6$, $3$, and $1.5\,\Mpc$, respectively.  For these
configurations, the characteristic U-shape of $Q_3(\alpha)$ is clearly
visible.  The 1-$\sigma$ errors are found to increase on larger scales
due to the reduced number of independent triangle configurations that
one can fit within the sample.  The S/N for these triangle
configurations are 8, 22, and 18 respectively, as indicated in the
bottom row of Fig.~\ref{fig:B_C_L_nl}. Such values are not
exceptional, but nevertheless are sufficient to allow useful
constraints on $B$ and $C$ to be determined.  These model constraints
are further discussed in section~\ref{sec:B_C_nl}.

\subsubsection{$Q_3(\alpha)$ as function of luminosity and colour} 
\label{sec:q3_L_nl} 
 
We use bright ($-21<$\Mbj$<-20$) and faint ($-19<$\Mbj$<-18$) volume
limited samples to investigate the luminosity dependence of
$Q_3(\alpha)$ in the non-linear regime. This is presented in
Fig.~\ref{fig:q3_L_nl} (top and bottom panels) for different triangle
configurations (from left to right). In Fig.~\ref{fig:B_C_L_nl}, we
show the corresponding likelihood contours for the model parameters
$B$ and $C$ from Eq.~\ref{eq:Q3G}. Comparing the results between the
faint and bright galaxy samples in Fig.~\ref{fig:q3_L_nl}, we see
that, even though the characteristic U-shape is generally preserved,
there is a weak tendency for the amplitude of $Q_3(\alpha)$ to
decrease with increasing galaxy luminosity. In terms of the model
parameters $B$ and $C$, Fig.~\ref{fig:B_C_L_nl} suggests that changing
the characteristic luminosity of the galaxy population results in a
shift in the best-fitting model.

\begin{figure} 
\centering{{\epsfxsize=0.47\textwidth \epsfbox{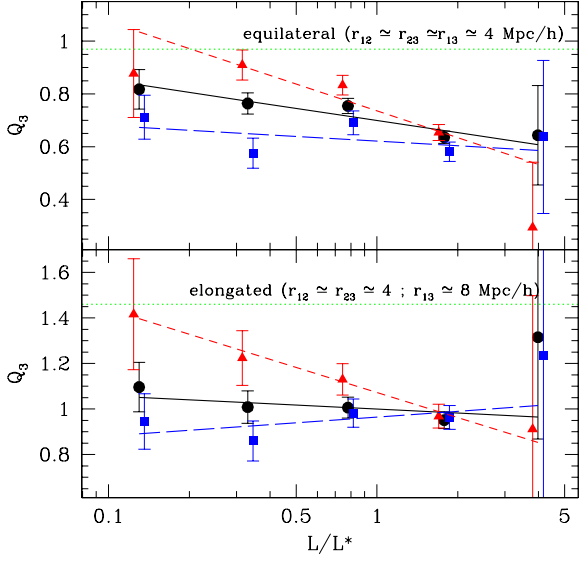}}}
\caption{Values of $Q_3$ for equilateral (upper panel) and elongated
  (lower panel) triangles, as a function of luminosity.  Measurements
  with errorbars show different galaxy samples: all galaxies
  (circles), red galaxies (triangles) and blue galaxies (squares).
  The solid, short-dashed and long-dashed lines show the corresponding
  best linear fit to Eq.~\ref{eq:Q3Lfit}, with the best-fitting values
  of the parameters quoted in Table~\ref{tab:q3Lfit}. The dotted
  horizontal line shows the corresponding \LCDM\ 
  prediction.  }
\label{fig:q3_L_colour}
\end{figure}

This general behaviour is better quantified in
Fig.~\ref{fig:q3_L_colour}, which shows how $Q_3$ changes with both
colour and luminosity.  The red and blue colour samples are sub-populations of
each volume limited sample, split by rest frame $b_{\rm J}-r_{\rm F}$
colour at $b_{\rm J}-r_{\rm F}=1.07$ (Cole \etal\ 2005). We focus our
attention on two characteristic configurations: equilateral triangles,
with $\alpha \simeq 60$ degrees, and elongated triangles with $\alpha
\simeq 180$. We choose a common scale around $r_{12} \simeq r_{23}
\simeq 4~\Mpc$, where all samples yield a good detection of $Q_3$.  To
improve the signal-to-noise, we take a large $\alpha$ bin, $\Delta
\alpha = \pm 18$ degrees, and a large $r_{12}$ bin, $\Delta r_{12} =
\pm 1~\Mpc$, compared to our standard choices of $\Delta \alpha = \pm
5$ degrees and $\Delta r_{12} = \pm 0.6~\Mpc$.
%% Indeed, from Fig.~\ref{fig:q3_scale} (discussed
%% in~\S\ref{sec:q3_scale}), no much scale dependence in $Q_3$ on the
%% mild non-linear regimes is detected.

The results in Fig.~\ref{fig:q3_L_colour} are well fit by a linear
relation that is a function of absolute magnitude:
\beq 
Q_3 = Q_3^* + \alpha_L~\log_{10}{L\over{L^*}} = Q_3^* - 0.4~\alpha_L~(M-M^*)~.
\label{eq:Q3Lfit}
\eeq
Table~\ref{tab:q3Lfit} shows the best-fitting values for the
parameters $Q_3^*$ and $\alpha_L$ using a $\chi^2$ fit with the full
covariance matrix.  As shown by the $\chi^2$ values in the table, the
functional form given in Eq.~\ref{eq:Q3Lfit} provides a very good
description of these measurements.  As seen in Table~\ref{tab:q3Lfit}
and also in Fig.~\ref{fig:q3_L_colour}, there is weak evidence for
luminosity segregation in both the blue and red
populations. Interestingly, the small but significant trend for the
overall population can be fully attributed to the luminosity
segregation in the red galaxies.

\begin{table} 
  \centering 
  \footnotesize 
  \caption[]{The best-fitting $Q_3^*$ and slope $\alpha_L$ of
    Eq.~\ref{eq:Q3Lfit} for different configurations (equilateral and
    elongated), for different galaxy populations (All, Red and Blue)
    and dark matter (DM).}
\label{tab:q3Lfit} 
  \begin{tabular}{ccccc}  
    \hline \hline 
    {Triangle} & {Sample} & $Q_3^*$  &  $ \alpha_L$ & $\chi^2$\\
    \hline
    equilateral & All  &  $0.70 \pm 0.02$ & $-0.15 \pm 0.04 $ & 3.1\\
    equilateral & Red  &  $0.74 \pm 0.04$ & $-0.34 \pm 0.11 $ & 4.2\\
    equilateral & Blue &  $0.62 \pm 0.03$ & $-0.06 \pm 0.07 $ & 4.3\\
    equilateral & DM &  $0.97 \pm 0.03 $ & \hphantom{0}0 &  \\
    \hline 
    elongated & All    &   $0.99 \pm 0.04$      &  $-0.06 \pm 0.09 $ & 1.7 \\ 
    elongated & Red    &   $1.07 \pm 0.02$      & $-0.37 \pm 0.03$  & 0.2\\
    elongated & Blue   &   $0.96 \pm 0.03$      & $\hphantom{00}0.08 \pm 0.08 $ & 1.2\\
    elongated & DM  &   $1.46 \pm 0.03 $  & \hphantom{0}0&  \\
\hline \hline 
  \end{tabular} 
\end{table}

\begin{figure} 
\centering{{\epsfxsize=0.47\textwidth \epsfbox{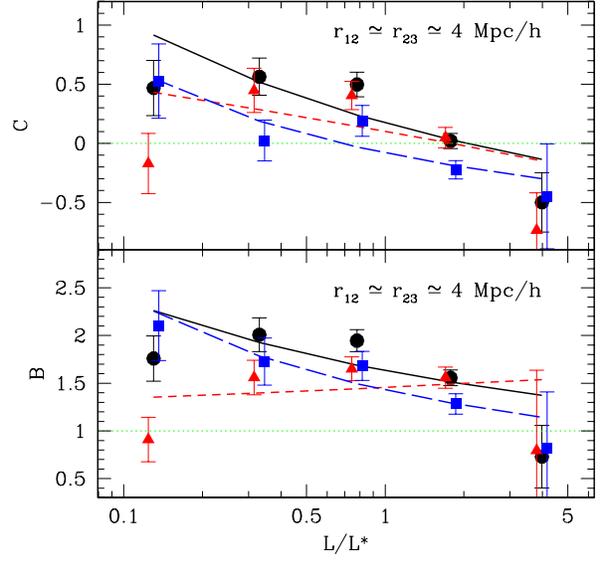}}}
\caption{Deviations of $Q_3$ from the \LCDM\ prediction as quantified
  by $B$ and $C$ (see Eq.~\ref{eq:Q3G}; recall that, for the dark
  matter, $B=1$ and $C=0$, as shown by the horizontal dotted line in
  each panel).  Results are shown for all configurations of triangles
  with $r_{12} \simeq r_{23} \simeq 4~\Mpc$, as a function of
  luminosity.  Circles, triangles and squares with errorbars
  correspond to all, red and blue galaxies respectively. Continuous,
  short-dashed and long-dashed lines show the corresponding
  predictions for all, red and blue galaxies using the best-fitting
  parameters listed in Table~\ref{tab:q3Lfit} and Eq.~\ref{eq:Q3Lfit}.
}
\label{fig:B_C_L_colour}
\end{figure} 

\subsubsection{Constraints on $B$ \& $C$ from non-linear scales} 
\label{sec:B_C_nl} 

Fig.~\ref{fig:B_C_L_colour} shows, as symbols with errorbars, the
best-fitting values for $B$ and $C$ (see Eq.~\ref{eq:Q3G}) in the
non-linear regime using {\it all} isosceles triangle configurations
with $r_{12} \simeq r_{23} \simeq 4~\Mpc$, i.e. not just the elongated
and equilateral cases considered in Fig.~\ref{fig:q3_L_colour}.  We
find weak systematic trends in the values of these parameters with
luminosity, with fainter galaxies favouring larger values of $B$ and
$C$. All samples, except the brightest, are at least 1 or 2-$\sigma$
away from the unbiased \LCDM\ case (i.e.  $B=1$ and $C=0$).
Fig.~\ref{fig:B_C_L_colour} also indicates, using lines, the
corresponding values of $B$ and $C$ when we use the best linear fit to
the $Q_3$ data as a function of luminosity, as listed in
Table~\ref{tab:q3Lfit}.  These lines emphasise the monotonic behaviour
of the data and provide some additional idea of the uncertainties.

We note that if one took the naive interpretation of $B$ as the linear
bias parameter, $b_1$, then the behaviour shown in
Fig.~\ref{fig:B_C_L_colour} is exactly the opposite to that previously
reported from the analysis of the 2-point correlation function,
$\xi_2$, over similar volume limited samples (Norberg \etal\ 2001,
2002a, 2005). However, in our case here, the measurement of the
parameter $B$ is done in the non-linear clustering regime, where $B$
\emph{can no longer be interpreted as the linear bias parameter}. This
is why our notation explicitly differentiates between $B$ and $b_1$,
and explains our choice to split the analysis into the two distinct
clustering regimes. On small scales, we expect strong corrections to
the linear relation $\xi_2 \simeq b_1^2 \xi_2^{DM}$ and similarly to
Eq.~\ref{eq:Q3G}. Indeed, the expansion used in Eq.~\ref{eq:bk} is
only valid in regimes where the density fluctuations are small with
$\delta < 1$.  Thus, the fact that $B$, as measured from
$Q_3(\alpha)$, decreases with increasing luminosity does not
necessarily mean that $\xi_2/\xi_2^{DM}$ should decrease for brighter
galaxies. Once we are in the non-linear regime, $B$ and $C$ can only
be understood in terms of their effect on $Q_3(\alpha)$, ie. a shift
and scale modification of the $Q_{3}^{\tt DM}(\alpha)$ dark matter
configuration, and not in terms of the theoretically motivated
relation given by Eq.~\ref{eq:bk}. Even with these interpretative
restrictions, the values of $B$ and $C$ that we recover provide an
accurate description of how biasing changes $Q_{3}^{\tt DM}(\alpha)$
and so give interesting new constraints on models of galaxy formation.

\begin{figure*} 
\centering{{\epsfxsize=0.47\textwidth \epsfbox{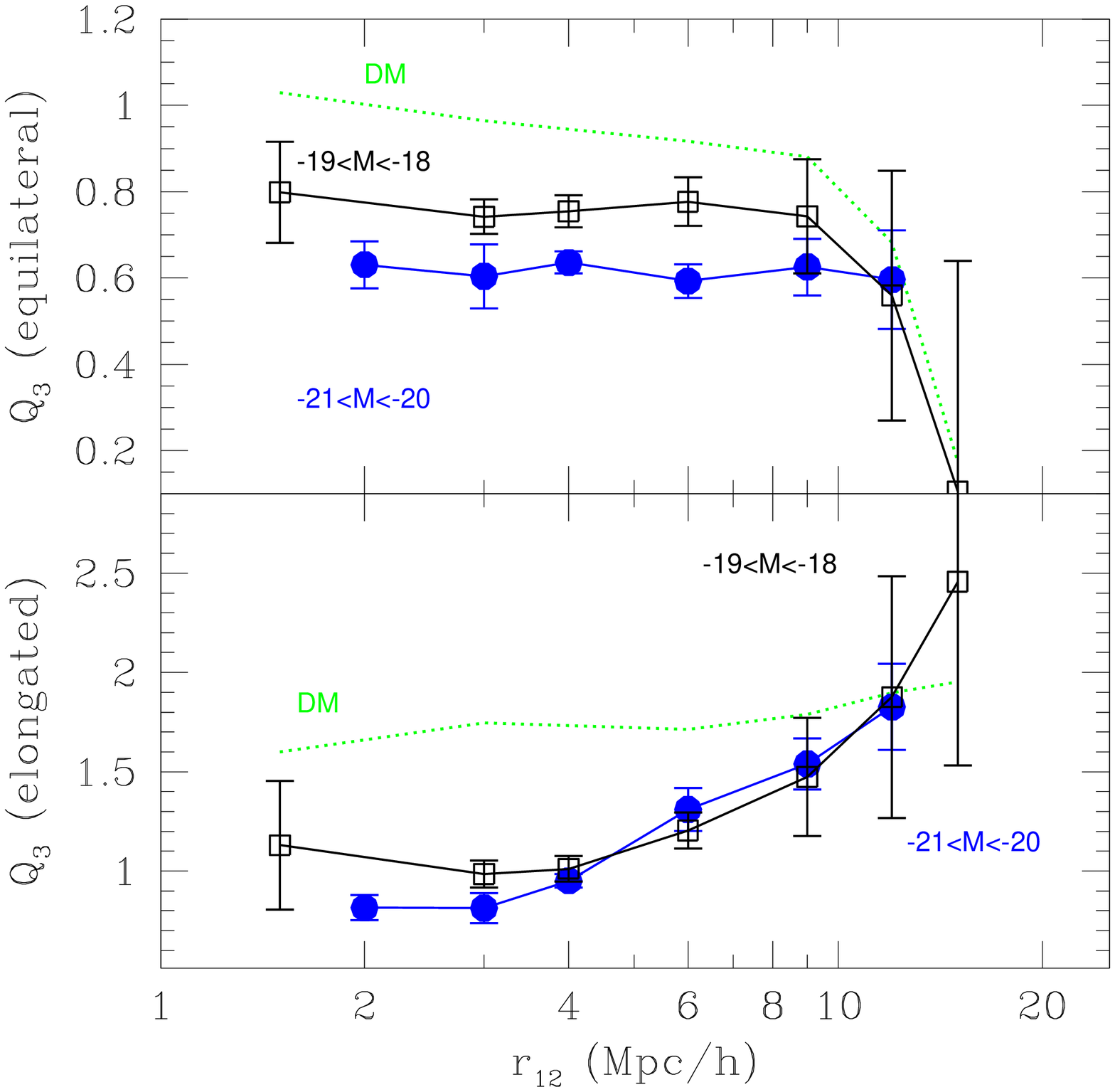}}
\epsfxsize=0.47\textwidth \epsfbox{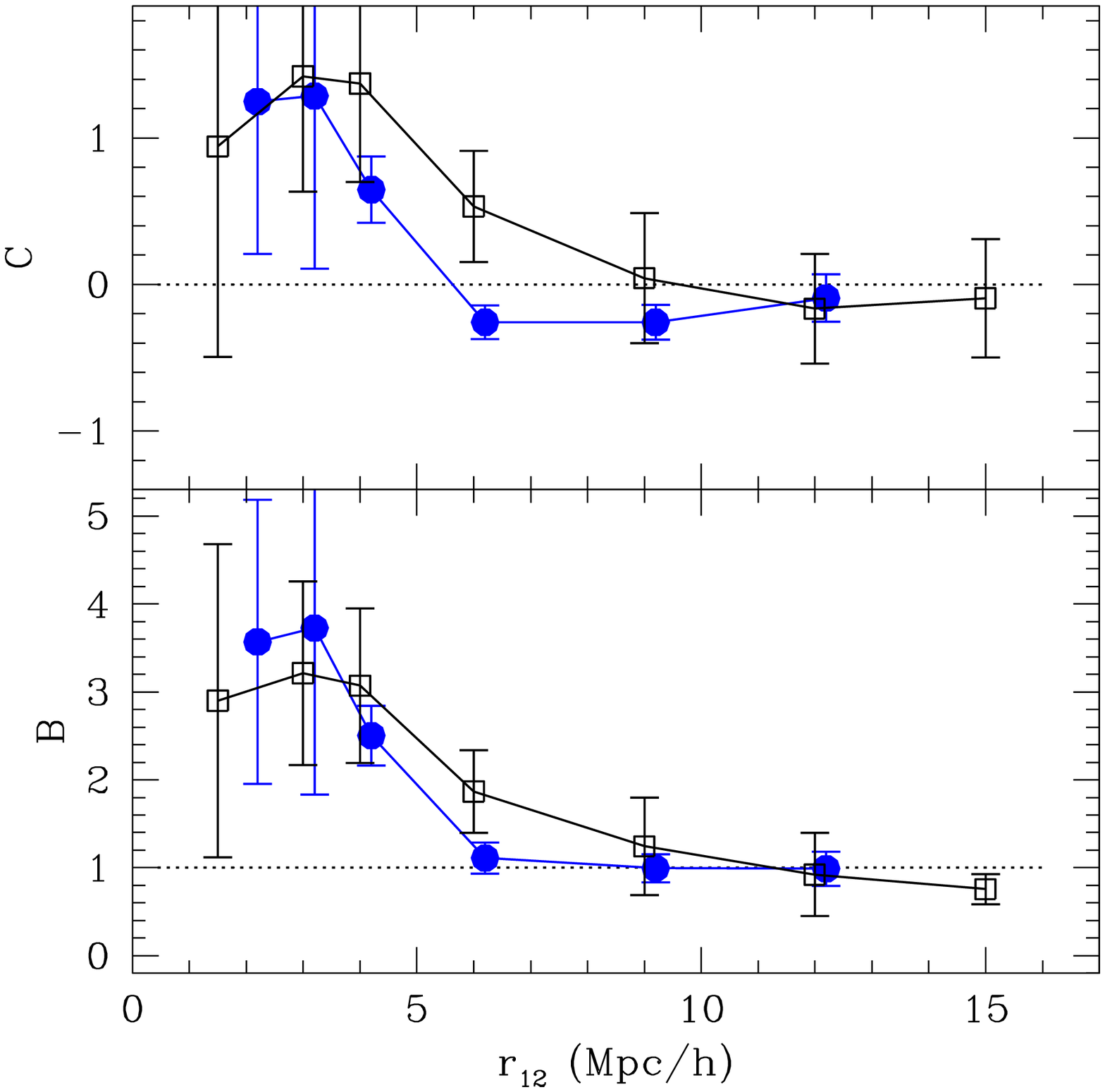}}
\caption{Left panels: the upper and lower panels show, as a function
  of the triangle side length, $Q_3$ for equilateral and elongated
  configurations respectively. Squares and dots correspond to the
  $-19<$\Mbj$<-18$ and $-21<$\Mbj$<-20$ galaxy samples
  respectively. For equilateral triangles, we detect a clear
  luminosity segregation on non-linear scales and no significant scale
  dependence for galaxies of a given luminosity.  This is in contrast
  to elongated triangles, where a strong scale dependence is observed
  but with no indication of luminosity segregation on any scale. The
  dotted lines in each panel show the \LCDM\ prediction
  for $Q_3(r_{12})$.  Right panels: Best-fitting $B$ and $C$ values
  for the combined configurations shown in the left panel, for the
  $-21<$\Mbj$<-20$ (dots) and $-19<$\Mbj$<-18$ (squares) samples. Note
  the change to a linear scale in the $x$-axis.}
\label{fig:q3_scale} 
\end{figure*}

\subsection{$Q_3$ as function of scale} 
\label{sec:q3_scale} 

We present in Fig.~\ref{fig:q3_scale} results for $Q_3$ using
equilateral and elongated triangle configurations for the
$-19<$\Mbj$<-18$ and $-21<$\Mbj$<-20$ volume limited samples.  These
two configurations correspond to isosceles triangles with $\alpha=60$
degrees and $\alpha=180$ degrees respectively.  The equilateral
configuration has the nice property that it can be fully characterized
by just one triangle side length, so that each triangle
samples a unique scale.  Fig.~\ref{fig:q3_scale} shows that for the
equilateral configuration (upper panel), $Q_3$ displays little scale
dependence, as opposed to the elongated case (lower panel), which
shows a rather strong scale dependence, much stronger than the dark
matter prediction for similar triangle configurations.  In addition,
on non-linear scales, the equilateral configuration also provides
clear evidence for luminosity segregation, although on weakly
non-linear scales the results are currently too uncertain to continue
making such a claim. This, unfortunately, mirrors our conclusions from
section~\ref{sec:q3_L_wnl}, using other triangle configurations.
Elongated measurements of $Q_{3}$, in contrast, display no luminosity
dependence at all.

From Fig.~\ref{fig:q3_scale}, we conclude that both
triangle configurations, when probing weakly non-linear scales, are in
reasonably good agreement with the \LCDM\ prediction,
favouring nevertheless a slightly negative value for $C$. This is in
good agreement with the results presented in Table~\ref{tab:q3_wnl}.
On non-linear scales, the difference between the data and the dark
matter prediction is striking and will provide a powerful constraints
on models of galaxy formation.
Note that the results presented in Fig.~\ref{fig:q3_scale} are fully 
consistent with those discussed in \S4.1.3. The aim of the exercise of 
considering isoceles triangles is to isolate the dependence of $Q_3$ 
on scale. To achieve this, the binning of $Q_3$ as a function 
of $\alpha$ is degraded and the range of values of $\alpha$ considered 
on weakly non-linear scales is reduced when compared with the earlier, 
less restrictive analysis in \S4.1.3 (which considered also triangles with 
$r_{13}>r_{12}$ and better $\alpha$ resolution). As a consequence, the 
resulting constraints on the parameters $B$ and $C$ presented in 
this subsection have larger errorbars. 

\begin{figure*} 
\centering{{\epsfxsize=0.47\textwidth \epsfbox{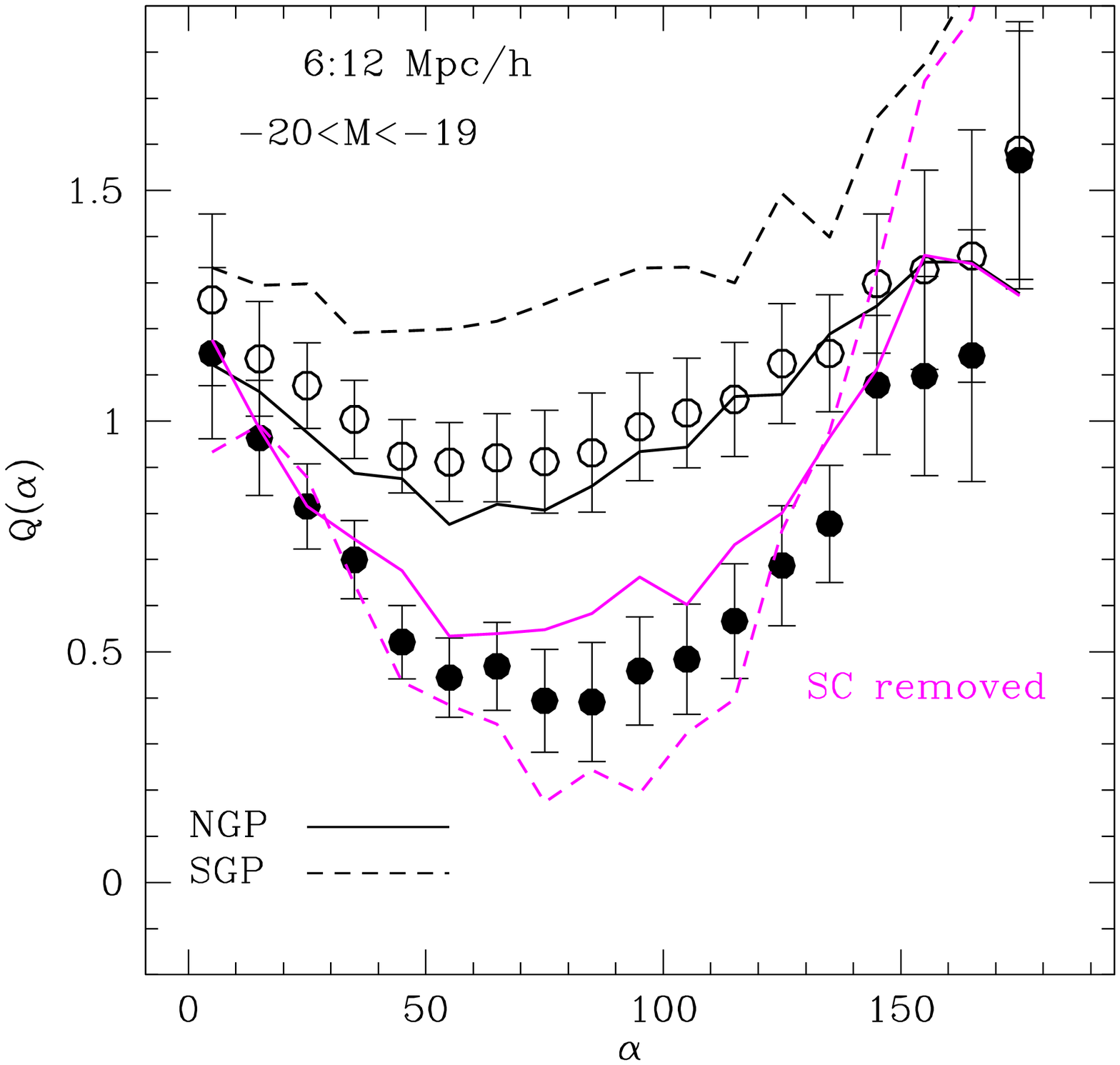}}
\epsfxsize=0.47\textwidth \epsfbox{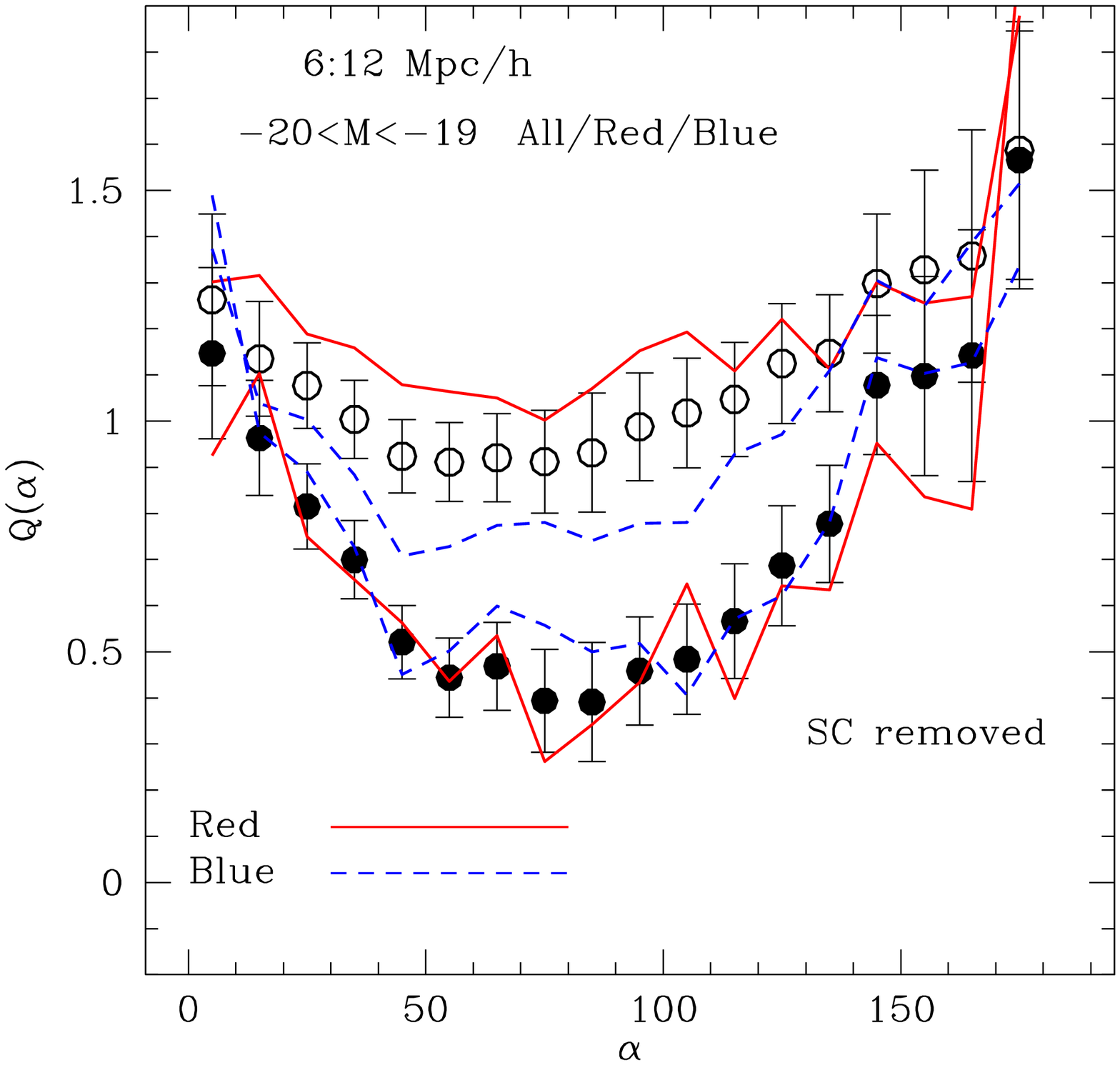}}
\caption{The impact of the superstructures on the measurement of $Q_3(\alpha)$
  for the $-20<$\Mbj$<-19$ sample and the $r_{13}=2~r_{12}=12~\Mpc$
  triangle configuration. Open and filled circles show respectively 
  the results with and without the superstructures (SC). All errorbars 
  are derived from the mocks and correspond to a 2-$\sigma$ limit.
  Left panel: solid (dashed) lines show results for the NGP (SGP) 
  region. The upper pair of lines and set of symbols show the results 
  with the SC; the lower set without SC. 
  Right panel: solid (dashed) lines correspond to blue (red) 
  galaxies. As in the left panel, upper pair of lines and set of 
  symbols show the results with the SC; the lower set without SC.}
\label{fig:q3_lstar} 
\end{figure*} 
 
\subsection{The influence of superstructures on $Q_3$} 
\label{sec:q3_lstar} 
 
The measured $Q_3$ for the \lstar\ sample, i.e. galaxies with absolute
magnitudes in the range $-20<$\Mbj$<-19$, can be strongly influenced
by the presence of two superstructures in the 2dFGRS, depending on the
scale measured. The impact of these structures on the distribution of
counts-in-cells in the \lstar\ sample was first pointed out by Baugh
\etal\ (2004).  Baugh \etal\ presented results both with and without
the superstructures, to illustrate the systematic effect that their
presence has upon the clustering measurements.  For the
\lstar\ sample, the superstructures were found to dominate the
clustering statistics on scales larger than $\approx 6~\Mpc$, and we
find that these structures have a comparable influence in our analysis, as shown in
Fig.~\ref{fig:q3_lstar} (in general the influence of these structures
can be seen out to the largest scales that can be probed). On smaller
scales ($r_{13}<6~\Mpc$), and for the other volume limited samples
that we consider, either the systematic contribution from these large
coherent structures lies within the (correlated) errorbars from the
mocks, or, because of their spatial location, the superstructures are
not present within the volume analysed.
 
The impact of the superstructures is equally pronounced in both the
NGP and SGP regions, as is clear from the left panel of
Fig.~\ref{fig:q3_lstar}. Both the NGP and SGP results for
$Q_3(\alpha)$ (solid and dashed lines respectively) change in a
similar way when the superstructure in each region is removed. As we
pointed out in Section~\ref{sec:2dfdata}, the volume masked out when
the superstructures are excised is less than $2~\%$ of the total, with
a loss of approximately $5~\%$ of the galaxies.  As Baugh
\etal\ remarked, the act of removing the superstructures is not
intended serve as a correction to the clustering measurements, but
rather as an illustration of the systematic effects that rare objects
produce in higher order clustering statistics. That said, the results
with the superclusters removed do appear to be more in line with our
theoretical prejudice for the weakly nonlinear regime.

The difference between the results obtained for $Q_3(\alpha)$ with and
without the superstructures is larger than the variance over our
ensemble of 22 mock catalogues.
%We note also that the variance  
%produced by the presence of these structures is larger than the variance found  
%from 22 mock galaxy samples, (comparison of middle left panel in Fig.\ref{fig:q3r3}) 
This indicates that the volume limited samples in the 22 mocks do not
contain the large, coherent structures seen in the real data, as
mentioned above.  As Baugh \etal\ (2004) commented, the presence of
such structures could give us new insights into models of structure
formation.  However, it is important to bear in mind that the lack of
superstructures in such a small number of mocks does not place a very
high confidence limit against them being seen at all in the
\LCDM\ model. All that we can conclude from the non-detection of such
objects in our ensemble of synthetic 2dFGRS catalogues is that they
occur less than 5\% of the time.  We have carried out an analysis of a
large ensemble of dark matter simulations, in addition to undertaking
a more extensive search of the Hubble Volume simulation, in order to
place tighter constraints on the frequency of such superstructures in
the \LCDM\ cosmology.  Before we can place firm limits on the chances
of finding superstructures like those seen in the 2dFGRS, we need to
make realistic mocks with the radial and angular selection of the
2dFGRS, but the preliminary indication from analyzing idealized,
cubical volumes is that {\it it is} possible to find such superstructures in
the simulations.  Full details will be presented in a future paper
(see also Fosalba, Pan \& Szapudi 2005, who discuss the impact
$~10^{15}$ solar masses halos on theoretical prediction of
$Q_3$).

In the right panel of Fig.~\ref{fig:q3_lstar}, we show, for the same
triangle configuration as in the left panel, the results for
$Q_3(\alpha)$ split by colour: red (solid) and blue (dashed). It is
interesting to see that with the superstructures included, the
difference between $Q_3$ measured for red and blue galaxies is mildly
significant (with red galaxies having a systematically larger $Q_3$).
In contrast, when the superclusters are excluded from the analysis,
the measured $Q_3$ for red and blue galaxies are identical to within
the errors. This segregation suggests that the superstructures are,
perhaps not unsurprisingly, populated preferentially by red galaxies.

\section{Comparison with previous 2\lowercase{d}FGRS results} 
\label{sec:comparison}

As mentioned in the introduction, Baugh et~al. (2004) and Croton et~al. (2004)
found the puzzling result that $S_3 \simeq 2$ for 2dFGRS galaxies 
in contrast to the theoretical value of $S_3^{DM}  \simeq 3$ expected 
in \LCDM\, on large (weakly non-linear) scales. 
This apparent inconsistencey can now be resolved using the bias parameters 
we have measured here, i.e. $b_1 \simeq 1$ and $c_2 \simeq -0.3$  gives
$S_3^G \simeq (S_3^{DM} +3 c_2)/b_1 \simeq 2$, in good agreement
with the above mesurements on large scales.
In their Fig.~10, Croton et~al. (2004) found a weak dependence of 
$S_3$ on galaxy luminosity, with a slope $B_3 \simeq -0.4$ detected 
with $2$-$\sigma$ confidence level. We note that this is in very good 
agreement  with our nearly $4$-$\sigma$ detection of 
$\alpha_L \simeq -0.15 \pm 0.04$, as quoted in Table~3. 
Since $S_3 \simeq 3 Q_3$, we would expect $B_3 \simeq 3 \alpha_L 
\simeq -0.45 \pm 0.12$, as found in Croton et~al. (2004), but 
with a higher significance.

Measurements of three-point statistics from early 2dFGRS data releases
were made by Verde et~al. (2002) and Jing \& B\"orner (2004).  These
authors used compilations comprising 127K and 100K galaxies
respectively. Here, as remarked earlier, we use the final dataset
which contains double the number of galaxies and double the volume
that were available for analysis in these preliminary studies.
Recently, Pan \& Szapudi (2005) have also analysed the final 2dFGRS
dataset, estimating the monopole moment of the three-point function,
averaging over the shape dependence of triangles. In this section, we
compare our results with those obtained by these authors and also with
the measurement of the projected $Q_3$ for the APM survey, the parent
catalogue of the 2dFGRS, made by Frieman \& Gazta\~naga (1999).
 
Verde et~al. (2002) found, using a three-point function analysis in
Fourier space, that 2dFGRS galaxies are essentially unbiased tracers
of the mass, recovering a linear bias factor consistent with unity,
$b_1=1.04 \pm 0.11$, and a second order bias that is effectively zero,
$b_2 \equiv b_1 c_2=-0.054 \pm 0.08$.  We note that Lahav
et~al. (2002) also reached a similar conclusion applying a different
approach to the same 2dFGRS dataset, arguing that $b_1 \sim 1$.  There
is good agreement between our best value for $b_1$ and that obtained
by Verde et~al, which is encouraging in view of the possible reasons
for discrepancies between the results of the two studies set out
below. However, our results for the quadratic bias are quite different
from those of Verde et~al. Our optimum measurement gives a 3-$\sigma$
detection of a non-zero value for the quadratic bias, whereas Verde
et~al. found a value consistent with zero.  The discrepancy between
our results and those of Verde et~al.  corresponds to $\Delta \chi^2
>80$ for 2 degrees of freedom.  This implies a 9-$\sigma$ discrepancy
(recalling that the errorboxes are not square, but elongated).  The
discrepancy in the claimed values of $c_2$ is only 3-4 -$\sigma$ if we
take the nominal errors on the measurement of $c_2$ by Verde
et~al. and assume square errorboxes.

What are the reasons behind this disagreement? We have identified some
aspects in which our analysis differs from that of Verde et~al., which
will contribute to the discrepancy at different levels, over and above
the fact that we used different versions of the 2dFGRS data.  First,
we have considered the full 2dFGRS in configuration space, thus
avoiding the need to compensate for the impact of the complicated
2dFGRS angular mask on measurements carried out in Fourier space.  
Verde et~al. do not correct for the convolution of the underlying
bispectrum with the angular survey window function, arguing that, for the
range of wavenumbers they consider, this effect is unimportant. 
This conclusion is based upon tests carried out for the 
{\it power spectrum} by Percival et~al.  (2001). 
The impact of the window function on the bispectrum could be more 
extensive than in the case of the power spectrum, introducing 
anisotropies into the recovered bispectrum, and has not 
been tested explicitly.  
Second, the range of galaxy luminosities considered is different in 
the two studies. We have analysed volume limited samples drawn 
from the 2dFGRS, whereas Verde et~al. used the flux limited 
survey (however our best measurement comes from galaxies with 
luminosities between 1.3 and 2.5~\lstar, and their sample corresponds 
to $\sim 1.9$~\lstar).  Third, the scales used to constrain the 
parameters of the bias model are also different. We use triangles 
that probe pair separations from 9 to $36~\Mpc$; Verde et~al. 
consider 13 to $62~\Mpc$, although most of their signal comes from 
the smaller scales, as shown by their fig.~4. Fourth, Verde et~al. 
neglect the covariance between measurements of the bispectrum at 
different wavenumbers, which is a poor approximation even in Fourier 
space, as shown by Scoccimarro et~al. (2001a) and Feldman et al (2001).  
Neglecting the covariance will artificially suppress the errors on 
$b_1$ and $b_2$ by a considerable factor, corresponding roughly to 
the actual number of bins used divided by the number of dominant 
eigenmodes of the reduced three-point function, which in this case 
could be up to a factor of 4. This could to some extent explain 
why our relative errors are larger than those quoted by Verde et~al., 
in spite of the more homogeneous 2dFGRS dataset used in our analysis. 
Verde et~al use mock catalogues to estimate the errors on the 
recovered values of $b_1$ and $b_2$. The true, underlying 
value of $b_2$ for the mocks is not known, so it is not possible 
to assess whether or not their method introduces any systematic biases 
in the recovered value of $b_2$. A bias on $b_2$ introduced by the 
convolution with the angular mask and the covariance in the bispectrum 
measurements could affect the estimated values of both the mean and 
the errors.  In fact, the mocks used by Verde et al. are 
very similar to the ones used here as they were produced using the
same prescription for galaxy biasing . As shown by the dashed lines 
in our Figs.~2,~3 and 5 there is a systematic shift of $Q_3$ in the
mocks with respect to the dark matter simulations, indicating 
that $b_2$ is in fact non zero (and negative) in the mocks, in
contrast to fig.2 of Verde et~al.
  
%Finally, Verde et~al. marginalize over the redshift
%space distortion parameter, $\beta$, and the pairwise velocity
%dispersion. Errors in the estimates of these parameters could have an
%impact on their results.  Verde et~al. used the values of these
%parameters obtained from the 2dFGRS by Peacock et~al. (2001).  In the
%more detailed analysis of the final 2dFGRS dataset carried out by
%Hawkins et~al. (2003), the errors quoted in the earlier study by
%Peacock et~al. are described as optimistic and these early results are
%barely within 1-$\sigma$ of the current best estimates.

Our results are in somewhat better agreement with those of Jing \&
B\"orner (2004) and Wang et al. (2004), who analysed the 2dFGRS 100k
release (Colless et~al. 2001). They found that $Q_3$ measured for the
2dFGRS is smaller than the $\Lambda$CDM predictions, particularly for
galaxies brighter than \lstar. This agrees with our result (compare
the measurements for galaxies shown by symbols in the top rows of
Figs.~\ref{fig:q3_210_wnl} and~\ref{fig:q3_L_wnl} with the dark matter
predictions plotted using thick lines) and is also at odds with the
Verde \etal\ result.  Our results for equilateral configurations in
Fig.\ref{fig:q3_L_colour} are also in good agreement with fig.~10 in
Wang et al. (2004).  However, the comparison with these results is not
straightforward for a number of reasons: (i) The authors used less
than half the data that we have analysed.  (ii) They used a different
parametrization and binning for their measurements of $Q_3$. (iii)
They neglected covariance between bins and used approximate bootstrap
errors.  Jing \& B\"orner interpreted the lower values of $Q_3$ that
they found as a consequence of a larger linear bias, $b_1 \simeq 1.5$,
in contrast to our conclusion that most of the bias comes from the
quadratic term $c_2 \simeq -0.3$, with a linear bias consistent with
unity.  This difference has a dramatic consequence for the implied
value of $\sigma_8$.  For galaxies fainter than \lstar , Jing \&
B\"orner get unbiased results, which disagrees with our
findings. This, however, could be a result of the smaller volume
probed by Jing \& B\"orner, which gives larger errors on their
measurement.  Jing \& B\"orner also seem to find less configuration
dependence for $Q_3$, i.e. as function of the triangle shape specified
by $\alpha$. As pointed out in GS05, this could partly be due to the
use of too large a bin in the $\alpha$ angle that parameterises
triangular shape in addition to the smaller volume used.

Most recently, as this paper was about to be submitted, Pan \& Szapudi
(2005) presented new results on the monopole moment of the three-point
function measured from the full 2dFGRS. They find $b_1 \simeq 1.04^{+0.23}_{-0.09}$ and $b_2 \simeq -0.06^{+0.03}_{-0.01}$. Both the
technique and assumptions employed by Pan \& Szapudi are conceptually
very different from ours.  The monopole contribution to the normalized
three point function merely yields a constant value that is
independent of triangle opening angle.  It is approximately equivalent
to the first eigenmode in our Singular Value Decomposition of the
covariance matrix of $Q_3$, and therefore contains much less
information than we use to place constraints on the bias parameters.
As a consequence, the monopole alone {\it cannot} be used to separate
$b_1$ from $b_2$; only the higher multipoles of $Q_3$ can break this
degeneracy.  Pan \& Szapudi instead use a simultaneous fit to the
amplitudes of the two ($\xi$) and three-point ($\zeta$) functions (as
a function of scale) to place separate constraints on the values of
$b_1$ and $b_2$; recall that our analysis only requires a fit to the
ratio $Q_3 \sim \zeta /\xi^2$.  Both the modelling and the systematics
involved in the fit used by Pan \& Szapudi are therefore quite
different from ours.  Our analysis is less sensitive to possible
systematics in the amplitude of $\zeta$. In particular, we do not need
to model the impact of redshift distortions on the amplitudes of the 2
and 3-point functions as Pan \& Szapudi must.
%The Pan \& Szapudi analysis ignores the shape dependence of $Q_3$ 
%and is therefore not sensitive to this systematic error.
Another important difference is the implicit assumption used by Pan \&
Szapudi that the biasing parameters, $b_i$, are constant over the
whole range of scales considered, i.e. from $4-60~\Mpc$.  In our case,
we allow $b_i$ to change for each combination of fixed scales $r_{12}$
and $r_{13}$.  Given these differences, there is surprisingly good
agreement in the values obtained for $b_1$ by the two methods.
However, their $b_2$ value is significantly different.  This is not
unexpected given the systematic uncertainties in modelling redshift
distortions through the $f_2^2$ term in eqn.~[6] of their paper.  As
shown in the right-hand panel of our Fig.~\ref{fig:q3_scale} and in
Table ~\ref{tab:q3_wnl}, we find a weak trend for the bias parameters
to increase as the triangle scale is reduced. This could also help to
explain the slightly larger biasing parameters they find. Perhaps more
puzzling is their fig.~7, which shows how $b_1$ increases for the
brightest galaxies, in contrast to our Fig.~7 (which only applies to
the smallest scales considered by Pan \& Szapudi).

Our findings are compatible with the values of the projected $Q_3$
measured in the APM survey, which is the parent catalogue of the
2dFGRS.  Frieman \& Gazta\~naga (1999) also found values of $Q_3$ that
lay below the $\Lambda$CDM predictions. However, they did not perform
a proper S/N analysis with the covariance matrix to separate $b_1$
from $b_2$.
\footnote{We note that such an analysis was, however, presented for
  $Q_3$ measured in Fourier space from the IRAS Point Source Redshift
  Catalogue (PSCz, Saunders et~al. 2000) by Scoccimarro \etal\ (2001a)
  and Feldman \etal\ (2001).}  Our results are also in qualitative
agreement with the values of the angular skewness, $s_3$, measured in
the APM galaxy survey (Gazta\~naga 1994). The mean over large angular
scales (corresponding to $7-30~\Mpc$) was estimated to be $s_3= 3.8$,
with an error (dominated by sampling covariance) of the order $\simgt
10~\%$.  The theoretical predictions for the projected moments using
perturbation theory (Bernardeau 1995) and dark matter simulations
(Gazta\~naga \& Bernardeau 1998) yield a mean value $s_3 \simeq 5$
over the same scales
\footnote{Note that this prediction differs from the hierarchical
  projection for the same model/scales estimated by Gazta\~naga
  (1994), which were closer to $s_3 \simeq 4$.  This was first noted
  by Bernardeau (1995) and later confirmed with simulations by
  Gazta\~naga \& Bernardeau (1998). See also comments relating to
  figs.~47 and 54 in Bernardeau et~al. (2002) for further details.}.
The skewness measured from the APM Survey is thus also below the the
\LCDM\ prediction.  This agrees well with our estimates of the bias
parameters $b_1 \simeq 1$ and $c_2 \simeq -0.3$, which give $s_3^G
\simeq (s_3+3c_2)/b_1 \simeq 4$, in excellent agreement with the
observed APM values. Note that the APM results correspond to
configuration space, in contrast to our results which are in
redshift space. Thus, our simple quadratic bias model (in redshift
space) can account  simultaneously for observations of real-space 
(projected) and redshift-space results for 3-point statistics 
(both skewness and 3-point function). Our new result solves the 
long standing observational puzzle regarding how the measured and 
predicted values of $S_3$ and $Q_3$ can be reconciled.

\section{Conclusions} 
\label{sec:con}

We have measured the reduced 3-point function $Q_3(r_1,r_2,r_3) \sim
\zeta /\xi^2$ (as defined in Eq.~\ref{fiftheq}) in the final 2dFGRS
catalogue, using triangles of different scales and opening angles. We
have utilized a range of volume limited samples in our analysis, which
allows us to look for clustering trends as a function of galaxy
luminosity. The inclusion of $r_{\rm F}$-band photometry in the final
2dFGRS data release also allows us to look for a dependence of the three point 
function on galaxy colour. Another novel aspect of our analysis is that we employ
an eigenmode decomposition to deal with correlations between
data-points and to assess the signal-to-noise of our measurements; our
results typically have a signal-to-noise $> 20$.

There are two primary motivations for measuring the reduced three
point function. The first is to test the gravitational instability
paradigm for the formation of large-scale structure in the Universe.
There are clear predictions for the form of the three-point function
in the case of an initially Gaussian distribution of density
fluctuations that have evolved under gravity (see Bernardeau
et~al. 2002).  The second motivation is to provide new constraints on
models of galaxy formation, by establishing how the three-point
function of galaxies differs from that of the underlying dark
matter. It turns out that the predictions for the dark matter are
insensitive to the amplitude of density fluctuations and to the
detailed shape of the power spectrum.

We have divided our analysis into two clustering regimes: weakly
non-linear clustering (i.e. $r \simgt 6 h^{-1}$Mpc or $\xi \simlt 1$)
and non-linear clustering ($r \simlt 6 h^{-1}$Mpc or $\xi \simgt 1$).
On weakly non-linear scales, there is a striking similarity between
the {\it shape} of $Q_3$ measured for galaxies and the predictions for
the dark matter.  This supports the idea that the basic phenomenon
behind the clustering pattern of galaxies is gravitational
instability, which confirms our previous conclusions reached from the
analysis of the distribution of counts-in-cells for the 2dFGRS (Baugh
et~al. 2004; Croton et~al. 2004b).

There are, however, significant differences, between $Q_3$ measured
for galaxies and the expectations for a \LCDM\ universe. We have
modelled this discrepancy in terms of a shift and a scaling applied to
the dark matter predictions. For scales on which the fluctuations are
weakly nonlinear, the scaling can be identified with the linear bias,
$b_1$ and the offset with the quadratic bias, $b_2/b_1$. Our best
measurement of these bias parameters gives a linear bias consistent
with unity, but a significant detection of a non-zero quadratic bias,
$b_2/b_1 = -0.34^{+0.11}_{-0.08}$.  This is the first time that the
signature of a quadratic bias has been seen so convincingly; our
measurements are 9-$\sigma$ away from the case in which galaxies
faithfully trace the mass ($b_1=1$ and $b_2=0$). Our results disagree
with some of the previous analyses of the three point function in the
2dFGRS; a detailed discussion of the possible reasons for this is
given in Section 5. We note that Feldman et~al. (2001) also found a
negative quadratic bias term when analysing the three point function
of galaxies in the IRAS Point Source Catalogue redshift survey, albeit
at a less significant level than our detection.

The discrepancy between $Q_3$ for galaxies and the dark matter
increases as the scale of the triangles is reduced (while remaining in
the weakly non-linear regime), which translates into a slight increase
in the best-fitting values of the bias parameters (see Table
~\ref{tab:q3_wnl}).  We find no significance evidence for luminosity
segregation on these weakly non-linear scales.

On smaller scales we are able to detect a significant dependence of
$Q_3$ on scale, color and luminosity. These trends appear at first
sight to be at odds with the preliminary results obtained by Kayo et
al. (2004) using the SDSS, although the errors on the measurements
presented by these authors are much larger than ours.  In all cases,
the measurements for the various samples of galaxies are clearly below
the predictions for the dark matter.  Our detailed measurements,
presented in Fig.6-8 and Table 3, should provide important new
constraints on models of galaxy formation (see Scoccimarro
\etal\ 2001b; Wang et al. 2004).

Our strong detection of a quadratic bias offers a new explanation of
the long standing puzzle of why redshift surveys have tended to
measure a different skewness ($S_3^{\tt G} \sim
\bar{\xi}_{3}/\bar{\xi}^{2}_{2}=2$; see Croton et~al. 2004b for the
2dFGRS and Table 19 of Bernardeau et~al. 2002 for a summary of other
observational results) from that predicted for the \LCDM\ cosmology
($S_3^{\tt M} \sim 3$).  If we take the non-linear bias relation
derived by Fry \& Gazta\~naga (1993), $S_3^{\tt G} = ( S_3^{\tt M} + 3
c_2)/b_1$ and insert our best-fitting values for the bias parameters
($b_1=0.95$ and $c_2=-0.35$), then we obtain $S3^G \simeq 2$, just as
required by the observations.

The value of $Q_3$ is independent of the overall amplitude of
fluctuations.  This means that our measurement of the linear bias,
$b_1$, is fully independent of the 
normalization of the fluctuations in the dark matter, as specified by
$\sigma_{8}$. Furthermore, the predictions for $Q_3$ for dark
matter are relatively insensitive to the shape of the power spectrum,
making this estimate of the bias robust to minor changes in the
parameters of the \LCDM\ model.  We can therefore combine our estimate
of $b_1$ with the amplitude of fluctuations measured from the galaxy
distribution, $\sigma^{\tt G}_{8}$, to derive an estimate of the
amplitude of fluctuations in the dark matter, $\sigma_{8}$.  Cole
et~al. (2005) measured the power spectrum of galaxy clustering in the
2dFGRS and found $\sigma_8^G \simeq 0.924 \pm 0.032$.  The equation
relating fluctuations in the galaxies to those in the dark matter
involves two other terms: 
\beq 
\sigma_8^{\tt G} = b_1~ D(z)~K(\beta)~\sigma_8~.  
\eeq 
Here $D(z) \simeq 0.95$ is the growth factor at the mean depth of the
survey ($z \simeq 0.1$) relative to the growth factor at $z=0$ and $K$
is the linear Kaiser (1987) redshift space distortion factor: $K
\simeq 1.17$ for $\beta \simeq 0.48$.  Both factors depend on the
cosmological density parameters for matter and vacuum energy, which we
have set to their concordance model values ($\Omega_m \simeq 0.3$,
$\Omega_\Lambda \simeq 0.7$ and $h \simeq 0.7$).  This allows us to
estimate $\sigma_8$: 
\beq 
\sigma_8 \simeq 0.88^{+0.12}_{-0.10}~.
\eeq 
Here we have assumed that the errors are dominated by the errors
in $b_1$.  This explains the good agreement found between the large
scale variance in 2dFGRS galaxies and the variance of the dark matter
for $\sigma_8 \simeq 0.9$, as shown in fig.~2 of Baugh \etal\ (2004).
A more detailed presentation of our result for $\sigma_8$ will be
deferred to a later paper.

{\bf Note added on submission:} 
On the day before our paper was submitted, Hikage et~al. (2005, 
astro-ph/0506194) posted a paper on the three-point function of SDSS 
galaxies, in which they perform a Fourier Phase analysis.
Their main result is that $b_{2}/b_{1} \approx 0$ if $\sigma_{8}=0.9$, in
apparent contradiction with our principal finding. However, Hikage
et~al. consider scales in excess of $30h^{-1}$Mpc and restrict their
attention to triangles with large opening angles. Their analysis is
therefore similar to the special case we present in Fig.~9 for elongated
and equilateral triangles. As we explained in \S4.3, in this case, due
to the reduced number of triangles considered, the errors on the bias
parameters are large. As shown in the upper right panel of our Fig.~2,
the error bars become quite large for $\alpha \simeq 180$ on large scales.
Furthermore, there is actually no reason to expect 
the bias parameters extracted by Hikage et~al. to agree closely with 
ours, as SDSS galaxies are red selected.

\section*{Acknowledgements} 
 
We acknowledge discussions with Roman Scoccimarro, Cristiano Porciani, 
Pablo Fosalba, Alan Heavens and Licia Verde. EG acknowledges
support from the Spanish Ministerio de Ciencia i Tecnologia, project
AYA2002-00850 with EC-FEDER funding. PN is supported by an ETH Zwicky
Fellowship. CMB is supported by a Royal Society University Research
Fellowship. DC acknowledges the financial support of the International
Max Planck Research School in Astrophysics Ph.D. fellowship.  This
work was supported by the EC's ALFA-II programme via its funding of
the Latin American European Network for Astrophysics and Cosmology.
The 2dFGRS was undertaken using the two-degree field spectrograph on
the Anglo-Australian Telescope, and we thank the 2dFGRS Team for its
release. The VLS and HV simulations in this paper were carried out by
the Virgo Supercomputing Consortium using computers based at the
Computing Centre of the Max-Planck Society in Garching and at the
Edinburgh parallel Computing Centre. The data are publicly available
at http://www.mpa-garching.mpg.de/NumCos .

\label{lastpage} 
 
\end{document}